\documentclass[a4paper,fleqn,usenatbib]{mnras}

\usepackage{newtxtext,newtxmath}

\usepackage[T1]{fontenc}
\usepackage{ae,aecompl}

\usepackage{graphicx}   
\usepackage{amsmath}    
\usepackage{amssymb}    
\usepackage{xcolor}
\usepackage{xspace}
\usepackage{symbols}

\usepackage{enumitem} 

\usepackage{etoolbox}
\makeatletter
\patchcmd\@combinedblfloats{\box\@outputbox}{\unvbox\@outputbox}{}{%
    \errmessage{\noexpand\@combinedblfloats could not be patched}%
}%
\makeatother

\title[Star formation in galaxy spheroids]{Heart of Darkness: the influence of galactic dynamics on quenching star formation in galaxy spheroids}

\author[J.\ Gensior, J.\ M.\ D.\ Kruijssen and B.\ W.\ Keller]{
Jindra Gensior,\thanks{E-mail: j.gensior@uni-heidelberg.de}
J. M. Diederik Kruijssen
and Benjamin W. Keller
\\
Astronomisches Rechen-Institut, Zentrum f{\"u}r Astronomie der Universit{\"a}t Heidelberg, M{\"o}nchhofstra{\ss}e 12-14, 69120 Heidelberg, Germany\\
}

\date{Accepted 2020 April 23. Received 2020 April 23; in original form 2020 February 4.}

\pubyear{2020}

\begin{document}
\label{firstpage}
\pagerange{\pageref{firstpage}--\pageref{lastpage}}
\maketitle

\begin{abstract}
Quenched galaxies are often observed to contain a strong bulge component. The key question is whether this reflects a causal connection -- can star formation be quenched dynamically by bulges or the spheroids of early-type galaxies? We systematically investigate the impact of these morphological components on star formation, by performing a suite of hydrodynamical simulations of isolated galaxies containing a spheroid. We vary the bulge mass and scale radius, while the total initial stellar, halo and gas mass are kept constant, with a gas fraction of 5~per~cent. In addition, we consider two different sub-grid star formation prescriptions. The first follows most simulations in the literature by assuming a constant star formation efficiency per free-fall time, whereas in the second model it depends on the gas virial parameter, following high-resolution simulations of turbulent fragmentation. Across all simulations, central spheroids increase the gas velocity dispersion towards the galactic centre. This increases the gravitational stability of the gas disc, suppresses fragmentation and star formation, and results in galaxies hosting extremely smooth and quiescent gas discs that fall below the galaxy main sequence. These effects amplify when using the more sophisticated, dynamics-dependent star formation model. Finally, we discover a pronounced relation between the central stellar surface density and star formation rate (SFR), such that the most bulge-dominated galaxies show the strongest deviation from the main sequence. We conclude that the SFR of galaxies is not only set by the balance between accretion and feedback, but carries a (sometimes dominant) dependence on the gravitational potential.
\end{abstract}

\begin{keywords}
galaxies: star formation -- galaxies: elliptical and lenticular, cD -- galaxies: ISM 
\end{keywords}



\section{Introduction} \label{s:intro}
It is a major open question how the physics of star formation on the scales of giant molecular clouds affect the macroscopic evolution of galaxies. Numerical simulations provide a controlled environment to test current hypotheses and identify the pertinent underlying physics. The main challenge for such experiments is that star formation takes place on length scales far below the resolution limit of modern high resolution galaxy and cosmological simulations \cite[e.g.][]{Hopkins2018b}. Therefore, it must be implemented as a sub-grid model, as pioneered by \cite{Katz1992, Cen1992}. The star formation rate (SFR) volume density can be expressed as
\begin{equation} \label{eq:SFR}
    \dot{\rho}_{\rm SFR} = \eff \frac{\rho}{\tff},
\end{equation}
where $\rho$ is the volume density of the star-forming gas and $\eff$ the star formation efficiency (SFE) per free-fall time, i.e.\ the fraction of gas that is converted to stars over a time-scale
\begin{equation} \label{eq:tff}
\tff = \sqrt{\frac{3\pi}{32G\rho}} .
\end{equation}
Because the free-fall time is proportional to $\rho^{-0.5}$, equation~(\ref{eq:SFR}) effectively represents a volumetric star formation relation of the \citet{Schmidt1959} and \citet{Kennicutt1998} form, $\dot{\rho}_{\rm SFR} \propto \rho^{1.5}$.

While the empirical scaling relation between SFR and gas surface density ($\Ssfr \propto \Sg^n$, with $n \simeq 1.4$, see e.g.\ \citealt{Kennicutt2012}) holds for late-type galaxies, which are traditionally considered to be star-forming \citep[e.g.][]{delosReyes2019}, there is increasing evidence that this relation does not hold universally. Molecular gas has been detected in about 22~per~cent of early-type galaxies \citep{Young2011, Davis2019}, with gas surface densities comparable to those found in late-types. However, galaxies with stellar spheroids\footnote{Here used interchangeably with `bulges' or `early-type galaxies' to indicate the presence of a spheroidal stellar morphological component.} systematically exhibit longer gas depletion times (the time it takes for the gas within a galaxy to be converted to stars at the current SFR) than late-type galaxies, with gas fractions being either lower than or comparable to those of their massive spiral counterparts \citep{Saintonge2012}. Similarly, analysing the SFRs of galaxies in the ATLAS$^{\rm3D}$ survey, \cite{Davis2014} showed that early-type galaxies exhibit lower SFRs compared to their late-type counterparts. \citet{Davis2014} find that the early-type star formation relation lies a factor of $\sim3$ below those of \cite{Kennicutt1998} and \cite{Bigiel2008}.

The low SFR in spheroids is not restricted to early-type galaxies. The Central Molecular Zone (CMZ, i.e.\ the central few 100~pc of the Milky Way) exhibits similar behaviour. Despite a large abundance of dense gas, the observed SFR in the CMZ falls below predictions of aforementioned empirical scaling relations \citep{Longmore2013,Kruijssen2014b}. The observation that star formation is suppressed across different (though all bulge-dominated) environments implies that physical processes beyond the simple density scaling of equation~(\ref{eq:SFR}) must be considered to fully understand star formation in galaxies.

`Morphological quenching' \citep{Martig2009} has been proposed as a phenomenological concept to explain the absence of star formation in the presence of molecular gas reservoirs. In this picture, galaxies with a dominant spheroidal component have a steeper gravitational potential well at the centre relative to disc-dominated galaxies. This means their angular velocity ($\Omega$) increases towards the bulge-dominated centre, raising the degree of gravitational stability (expressed through the $Q$ parameter of \citealt{Toomre1964}). \citet{Martig2009} speculate that shear, caused by a larger and more peaked $\Omega$ profile, can induce and maintain a high turbulent velocity dispersion ($\sigma$) in the interstellar medium (ISM). This would then allow the gas to remain pressure-supported against collapse, thereby quenching star formation in the spheroid.

In the decade since being proposed, morphological quenching has been a mixed success. Observationally, \cite{Huang2014} identify a strong correlation between gas depletion time and specific SFR (sSFR, the SFR normalised by the stellar mass). However, only a weak link between stellar density and star formation has been found, implying that while stellar bulges and low SFRs might correlate, they need not be causally related. While the green valley galaxies\footnote{In sSFR--stellar mass space, these fall in between the `main sequence' of star-forming galaxies and the quenched galaxy population, and are therefore thought to be in the process of being quenched.} studied in \cite{Belfiore2018} exhibit a strong suppression of sSFR in the galaxy centres, they also show an integrated (but smaller) suppression in sSFR compared to galaxies on the `main sequence' of star-forming galaxies. This global suppression is restricted to galaxies with both a high central stellar density and a low-ionization emission line region. Some mass-matched galaxies with similarly high central densities exhibit `normal' star formation activity, thus suggesting that morphological quenching can not be the sole driver of suppressed star formation \citep{Belfiore2018}. By contrast, \cite{MendezAbreu2019} find a clear difference in star formation relation for bulge and disc components, with bulges exhibiting systematically lower SFRs and longer gas depletion times. Similarly to the results of \cite{Belfiore2018}, the sSFR profiles for their bulge components also exhibit a strong central and general suppression, suggesting that a dominant bulge can affect the star formation activity at all radii. As the gas content in bulge and disc components of their galaxies is similar, \cite{MendezAbreu2019} argue that the observed suppression in star formation is caused by a dynamical process that stabilises the gas, such as morphological quenching.

The numerical perspective paints a similarly ambiguous picture. In high-resolution simulations comparing an isolated spiral and elliptical galaxy, \cite{Martig2013} found that morphological quenching is effective, albeit only at gas fractions lower than a few per cent. Even for gas fractions as low as 4.5~per~cent, these authors find that the star formation relation for the two galaxy types is offset by only $\sim30$~per~cent. Investigating a variety of quenching mechanisms in isolated galaxy simulations, \cite{Su2019} only find a slight decrease of the SFR in their bulge-dominated galaxies, concluding that other quenching mechanisms are required to significantly suppress the star formation in these objects. Recently, \cite{Kretschmer2019} have demonstrated that morphological quenching can be reproduced in a cosmological zoom-in simulation when modelling star formation with a dependence on the virial parameter and Mach number of the gas, but not when using a constant SFE per free-fall time ($\eff$ in equation~\ref{eq:SFR}).

In this paper, we use a suite of hydrodynamical simulations of isolated galaxies to systematically quantify the effect of a spheroidal stellar component on the structure and dynamics of the ISM and galactic star formation, paying particular attention to the extent to which star formation can be suppressed by the presence of this component. As recent studies have shown, a physically-motivated, environmentally-dependent star formation model (captured through the choice of $\eff$) may be crucial for reproducing the observed effect. We therefore consider two different sub-grid prescriptions for star formation in this work. Throughout, we adopt a gas fraction of 5~per~cent, appropriate for Milky Way-mass galaxies at $z=0$, as well as for galaxies in the transition region between the main sequence and the quenched galaxy population \citep[e.g.][]{Saintonge2017,Catinella2018}. The paper is structured as follows. Section~\ref{s:SFiGS} presents a discussion of our sub-grid star formation prescription, discussing how star formation is traditionally modelled, before introducing our new sub-grid model that accounts for the dynamical state of the gas. The simulation suite is then introduced in Section~\ref{s:Meth}. To investigate the effect of the star formation model, Section~\ref{s:cvSFE} compares simulations using our dynamics-dependent model to those using a constant efficiency per free fall time $\eff$, while Section~\ref{s:BC} focuses on the impact of the bulge component on the galaxy. Section~\ref{s:Dis} places our results in context of recent observations and theoretical studies in the literature. Finally, we summarise our findings and conclude in Section~\ref{s:SC}.

\section{Modelling Star Formation in galaxy simulations} \label{s:SFiGS}
\subsection{Star Formation in the literature} \label{ss:SFiL}

Simply using equation~\ref{eq:SFR} as sub-grid star formation model theoretically allows star formation everywhere, regardless of the physical or dynamical properties of the ISM (even if the rate of star formation depends on the gas density). As star formation is observed to proceed in cold, dense, molecular gas \citep[e.g.][]{Wong2002, Bigiel2008}, a variety of thresholds are used in combination with the sub-grid model to prevent arbitrary and spurious star formation in e.g.\ hot, diffuse gas. A volume density threshold is the most common restriction used \citep[e.g.][]{Navarro1993, Springel2003, Kim2016}. Only gas denser than the threshold can form stars, which is often used as analogous to the overdensities that are Jeans unstable and will eventually collapse. In addition, some models ensure that stars only form under the appropriate physical conditions by imposing a temperature ceiling for star formation \citep[e.g.][]{Stinson2006, Nickerson2019} or incorporating an H$_2$ fraction into equation~\ref{eq:SFR} \citep[e.g.][]{Robertson2008, Christensen2012, Grisdale2017}.

The other main free parameter in star formation models defined by equation~(\ref{eq:SFR}) is the efficiency per free-fall time $\eff$. Both $\eff$ and the thresholds mentioned above are generally chosen such that the star formation in the simulated galaxy matches the observed relationship between SFR and gas surface density \citep{Kennicutt1998, Bigiel2008, Leroy2013}. However, recent work suggests that the SFR is set by the balance between inflow and feedback-driven outflow, such that these self-regulate and the star formation prescription itself only has a weak effect on the SFR \citep{Hopkins2011, Agertz2013}. This weakens the importance of the choice of thresholds and $\eff$. Strong and efficient feedback will shape the density structure and kinematics of the ISM, and thus inhibit star formation by preventing gas from simply collapsing into dense peaks through gravity \citep{Hopkins2011}. The SFE itself determines how many stars form and subsequently how much feedback is injected into the gas. \cite{Agertz2013} demonstrate that, when including efficient momentum input from stellar feedback, changing $\eff$ by a factor of 10 between simulations leads to a star formation relation that varies by a factor of 2 at most. This results in an effective degeneracy that is reflected by a wide range across the literature of (density) thresholds (these are somewhat resolution-dependent, but range from 0.1~$\ccm$ in \citealt{Pillepich2018} to 1000~$\ccm$ in \citealt{Hopkins2018b}) and efficiencies ($\eff=0.01$ in \citealt{Kim2016} to $\eff=1$ for the FIRE2 simulations of \citealt{Hopkins2018b}, even if the latter include additional criteria depending on the state of the gas, including self-gravity).

The common underlying assumption in the aforementioned simulations is that it is accurate to convert gas to stars with a constant $\eff$ once the criteria for star formation are satisfied. However, there is observational evidence \citep[and references therein]{Utomo2018, Schruba2019} that the SFE varies up to an order of magnitude both within and between galaxies. Similarly, \citet{Chevance2020} find that the galactic environment (and galactic dynamics in particular) often determine the lifetimes of molecular clouds and thus their integrated star formation efficiencies. Analytical studies \citep{Krumholz2005, Hennebelle2011, Padoan2011, Federrath2012, Burkhart2018} corroborate this further, all predicting an additional dependence of the SFR on the turbulent state of the ISM, rather than just gas self-gravity. All of these studies conclude that the SFE depends on the local environmental conditions. In view of our goal to assess how gas dynamics in stellar spheroids can impact (galactic-scale) star formation, it is crucial to include some form of environmental dependence in the star formation modelling of our simulations. 

Most theoretical predictions for $\eff$ are based on analysis of the gas density probability distribution function (PDF) and link it to the virial parameter $\avir$, the sonic Mach number $\mathcal{M}$, the turbulent forcing parameter $b$ and the ratio of thermal to magnetic pressure, $\beta$ \citep{Federrath2012}. When investigating these dependences in high resolution simulations of turbulent molecular clouds, \citet{Padoan2012, Padoan2017} show that of these four parameters, the SFE per free-fall time primarily depends on the virial parameter. Recently, these findings have been used by \cite{Semenov2016} and \citet{Kretschmer2019} as a sub-grid star formation model, in combination with a sub-grid model for gas turbulence, as well as by e.g.\  \citet{Kimm2017}, \citet{Trebitsch2017} and \citet{Rosdahl2018} without including a sub-grid turbulence model. Similar to the work done by these groups, we have developed a sub-grid star formation model based on the gas dynamics for use in the moving mesh code {\sc Arepo} \citep{Springel2010}, which we introduce next. 

\subsection{A sub-grid star formation model based on gas dynamics} \label{ss:SFSGM}
In order to account for the dependence of $\eff$ on the virial parameter found in numerical simulations of turbulent fragmentation, we implement an environmentally-dependent $\eff$ based on the parametrisation of \cite{Padoan2017}. It directly expresses $\eff$ in terms of the virial parameter of the gas, by writing
\begin{equation} \label{eq:eff}
    \eff = 0.4 \exp{ \left(-1.6 \avir^{0.5}\right)} .
\end{equation}
The virial parameter is approximately the ratio of turbulent to gravitational potential energy of a (molecular) gas cloud and can also be expressed as ratio of free-fall to turbulent crossing time ($\tcr$) of the cloud \citep{Bertoldi1992}, i.e.\
\begin{equation} \label{eq:avir}
  \avir = \frac{40}{3\pi^2}\left(\frac{\tff}{\tcr}\right)^2 , 
\end{equation}
with 
\begin{equation} \label{eq:tcr}
  \tcr = \frac{L}{2\sigma} ,
\end{equation}
where $L$ is the length scale associated with turbulence, over which $\sigma$ and $\avir$ are calculated. Substituting equations~(\ref{eq:tff}), (\ref{eq:avir}) and (\ref{eq:tcr}) into equation~(\ref{eq:eff}) yields the final expression for the SFE per free-fall time for each gas cell:
\begin{equation} \label{eq:eff_final}
  \eff \simeq 0.4 \exp{\left(- \frac{2.018}{\sqrt{G}} \frac{\sigma}{L \rho^{0.5}}\right)}.
\end{equation}

For any numerical implementation of this model, it is necessary to define the length scale $L$. Without a sub-grid model for turbulence such as in \cite{Semenov2016}, which enables these authors to associate $L$ with the size of a resolution element, we must decide on a suitable length scale that allows the model to work independently of resolution. As the virial parameter is a cloud-scale property, we need to evaluate it for local overdensities. To achieve this, we use a version of Sobolev's approximation \citep{Sobolev1960} to determine the size of a local overdensity around a gas cell. In analogy with Sobolev's original approximation, which uses velocity gradients to determine the characteristic size scale for radiative transfer in stellar envelopes, we define $L$ based on the characteristic length scale for changes in the density of the surrounding gas as set by the density gradient $|\dg| = |\rm d\rho/\rm d r|$, i.e.\
\begin{equation}\label{eq:Lcr}
  L = \left| \frac{\rho}{\dg} \right| ,
\end{equation}
which we refer to as density gradient length scale. 

This now shifts the focus onto how to calculate $\dg$. Because {\sc Arepo} is an Eulerian-Lagrangian hybrid method, one could obtain the gradient either using the cell interfaces of the mesh, or using a smoothing kernel like a smooth particle hydrodynamics (SPH) code. Using only the neighbouring cell interfaces introduces a resolution dependence, which is undesirable. Additionally, the virial parameter calculation would not be self-consistent, as $\sigma$, $\rho$ and $\dg$ would all be calculated on different length scales. Calculating the density gradient using a smoothing kernel ($W$) instead causes a dependence on the number of neighbouring cells ($n_{\rm ngb}$) picked up by the kernel and in turn on the smoothing length $h$, as $h$ is traditionally chosen such that $n_{\rm ngb}$ is a fixed, pre-determined number. As with using cell interfaces, this would make $\dg$ dependent on a fixed number of neighbours, and cause us to evaluate $\sigma$, $\dg$ and $\rho$ on different scales. Instead, we therefore evaluate $\dg$ and $\rho$ on a length scale which matches the density gradient length scale. To achieve this, we use the kernel approach of equation~(\ref{eq:densgrad_SPH}):
\begin{equation} \label{eq:densgrad_SPH}
    \nabla \rho(\mathbf{r}) = \sum_{j=1}^{n_{\rm ngb}} m_j \nabla W(|\mathbf{r}-\mathbf{r_j}|,h) . 
\end{equation}
However, following the above reasoning, we consider the number of neighbours an independent variable that is decoupled from the smoothing length. We keep $h$ fixed and introduce a third length scale $\ltw$, which is the distance from the central gas cell on which we find neighbours. This allows us to vary $\ltw$ in an iterative process aimed at matching the density gradient length scale $L$ (calculated based on $\dg$ and $\rho$) and $\ltw$.

\begin{figure*}
    \centering
    \includegraphics[width=1.\linewidth]{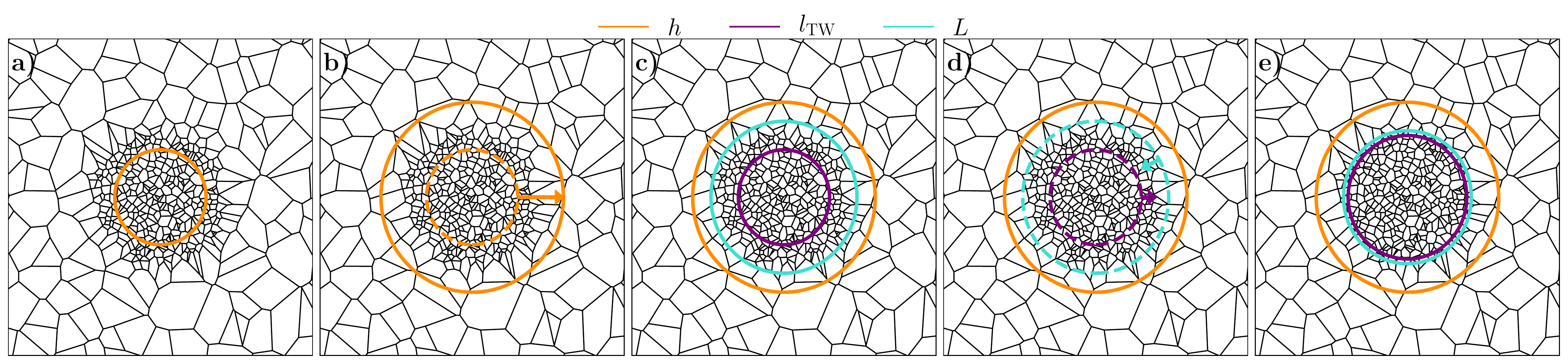}%
    \caption{Idealised overdensity on a Voronoi mesh; coloured lines indicate the smoothing length of the kernel ($h$, orange), the tree-walk length scale ($\ltw$, purple) and the density gradient length scale ($L$, cyan). 
    The five panels illustrate our algorithm for calculating the density gradient length scale (and subsequently the virial parameter).
    Panels a and b show how $h$ is set by identifying a distance within which a certain number of neighbours can be found (panel a) and then extending it by a factor of 2. After that, we keep $h$ constant, as seen in panels c--e. 
    We then find cells within a distance $\ltw$ and calculate $L$ based on their density and density gradient (panel c). Panel d shows how $\ltw$ is adjusted and iterated over, because $\ltw$ and $L$ do not match for the initial $\ltw$ in panel c. The overdensity is successfully identified once $\ltw$ and $L$ converge (panel e).
    }
    \label{fig:code_schematic}
\end{figure*}

Figure~\ref{fig:code_schematic} visualises how the code works and how the length scales are determined for an overdensity. Specifically, it progresses as follows:
\begin{enumerate}[label*=\arabic*,leftmargin=0.25cm]
    \item Determine and fix $h$:
    \begin{enumerate}[label*=.\arabic*,leftmargin=0.4cm]
        \item Perform a tree-walk to find 32 weighted neighbours (Figure~\ref{fig:code_schematic}, panel a)
        \item Set $h$ to $2\times$ the distance to the furthest neighbour (Figure~\ref{fig:code_schematic}, panel b), to not be too limited by the choice of $n$th neighbour and the distance to it
    \end{enumerate}
    \item Iterate over $\ltw$, at fixed $h$, until $|\ltw - L| \leq c_{\rm tw}$ is fulfilled (Figure~\ref{fig:code_schematic}, panels c--e)
    \begin{enumerate}[label*=.\arabic*,leftmargin=0.5cm]
        \item Our initial guess for $\ltw$ is set as $\ltw = 0.5h$ (see panel c).
        \item Adjust $\ltw$ by multiplying/dividing by a small factor (depending on the change in density gradient, but capped at $\sqrt{2}$) to become closer to the current value of $L$, for $L>\ltw$ (panel d) and $L<\ltw$ respectively; repeat until convergence is reached.  
    \end{enumerate}
\end{enumerate}
Once the density gradient and tree-walk length scales match to within the convergence criteria, we evaluate the velocity dispersion over the gas cell neighbours within $\ltw$ and include the thermal component from the sound speed $\cs$ when calculating $\avir$, i.e.\
\begin{equation}\label{eq:sigma_true}
  \sigma = \sqrt{\sigma_{\rm gas}^2 + \cs^2}.
\end{equation}
Now all quantities relevant for $\eff$ are known and the star formation rate can be calculated. This self-consistent approach distinguishes our sub-grid star formation model from other models in the literature that also do not include a sub-grid prescription for turbulence, in which the velocity dispersion is calculated from the velocity gradient across the star forming cell using only the nearest neighbours \citep[e.g.][]{Kimm2017}. We refer to Appendix~\ref{app:ResTest} for a quantitative demonstration of how the sub-grid model is only weakly affected by resolution.

\section{Method} \label{s:Meth}
The simulations described in this paper have been run with the moving-mesh code {\sc Arepo} \citep{Springel2010}. The equations of hydrodynamics are solved on an unstructured mesh, built from a Voronoi tesselation, using a second-order accurate, unsplit Godunov solver. The hydrodynamics solution is Galilean invariant, because the Voronoi generator points move with the gas fluid. Collisionless particles (i.e.\ stars and dark matter) are treated as Langrangian, with gravity being solved using a tree-based scheme. To achieve optimal gravitational resolution, we use an adaptive gravitational softening length \citep{Price2007}.

\subsection{Star Formation, Feedback and Cooling}

\subsubsection{Star Formation and Cooling} \label{sss:SFaC}
To model the thermal state of the ISM, we use the Grackle chemistry and cooling library\footnote{https://grackle.readthedocs.io/} \citep{Smith2017}, with the 6 species non-equilibrium chemistry network. This means we track atomic hydrogen, helium and their ions throughout the simulation. In combination with tabulated metal abundances these are then used to determine the cooling rate. Interstellar radiation is taken into account by including the \cite{Haardt2012} constant UV-background. As the galaxies in this study resemble evolved objects, we assume solar metallicity for each galaxy. 

The SFR of a gas cell is calculated as described in section \ref{ss:SFSGM}, using a cubic spline kernel \citep{Monaghan1992}.\footnote{To isolate the effect of the new sub-grid star formation model from any changes caused by the differences in gravitational potential, we repeat three of our simulations with a constant $\eff = 1$~per~cent (see Section~\ref{ss:ICs}).} We require the length scales $L$ and $\ltw$ to agree to within 10~per~cent of the smoothing length during the velocity dispersion and density gradient calculation, i.e.\ $c_{\rm tw} = 0.1 h$. Whether a gas cell is converted to, or spawns a star particle is then decided stochastically.

As discussed in section~\ref{ss:SFiL}, further constraints on the properties of star-forming gas are required to prevent spurious star formation, especially in simulations with a constant $\eff$ (see Section~\ref{ss:ICs}). Based on the values used in the literature and appropriate for our resolution, we use a minimum density threshold of $1~\ccm$, as well as a maximum temperature threshold of $10^3$~K. To enable a fair comparison between the constant and dynamics-dependent SFE, we apply the same restrictions to gas in simulations run with the virial parameter-dependent $\eff$. The dynamics-dependent model does not depend on these thresholds as strongly as the model with a constant SFE, because $\eff$ is regulated by the state of the gas. We refer to Appendix~\ref{app:SFC} for a more detailed discussion of how the results obtained with the dynamics-dependent SFE are only weakly affected by the choice of thresholds. 

\subsubsection{Feedback model} \label{sss:FB}
To study the effects of the star formation model on a gas disc, we must include a model for stellar feedback.  In the controlled experiments presented here, we use a simple model for feedback from Type II supernovae (SNe), first introduced by \citet{Kimm2014,Hopkins2014}. This `mechanical feedback' has excellent numerical convergence properties, with the same amount of total momentum injected over 6 decades of mass resolution \citep{Hopkins2018a}.  As was shown in \citet{Rosdahl2017}, this model also produces similar self-regulating behaviour to the stochastic thermal model of \citet{DallaVecchia2012} and the kinetic model of \citet{Dubois2008}, both of which are widely used in both cosmological and isolated simulations of galaxy formation.

We follow an approach similar to \citet{Hopkins2014}, where a kernel as in SPH is used to deposit feedback to the 32 nearest neighbours. These receive a share of feedback mass, metals, momentum, and thermal energy as a function of their M4 kernel weighting $W_{ij}$. For a given total SNe mass $M_{ej}$ and energy $E_{51} = E_{\rm SN}/10^{51}\ergs$, we calculate the terminal momentum of the blastwave at the end of the pressure-driven snowplough phase (and the beginning of the momentum-conserving snowplough phase) following \citet[equation 4.7]{Cioffi1988}, i.e.\
\begin{equation}
    p_{\rm term} = 4.8\times10^5~\Msun~\kms~
    \frac{(W_{ij} E_{51})^{13/14}}{\zeta_m^{3/14}n_0^{1/7}} ,
\end{equation}
with the metallicity parameter $\zeta_m = {\rm MIN}(Z/Z_\odot,0.01)$ and the gas density $n_0$. For each resolution element receiving feedback, we calculate an energy conserving (Sedov-Taylor) momentum $p_{\rm ST} = \sqrt{2W_{ij}m_{i}E_{\rm SN}}$ (where $m_i$ is the element mass after receiving ejecta) as well as the terminal snowplough momentum $p_{\rm term}$. Each element then receives feedback momentum $p_{\rm fb} = {\rm MIN}(W_{ij}p_{\rm term}, p_{\rm ST})$ and thermal energy $E_{\rm SN} = W_{ij}E_{\rm SN}-2(p-p_{\rm fb})^2/m_i$ (i.e.\ what remains).  

We assume a canonical $E_{51}=1$ for the SNe energy, with one SNe occurring per $100~\Msun$ of stellar mass formed \citep{Chabrier2003,Leitherer2014} and a delay time of $4~\Myr$ before SNe detonate. For simplicity, we detonate all SNe together $4~\Myr$ after a star particle has formed, which \citet{Kimm2015} showed to have little change in the overall star formation history by $z\sim3$ compared to having individual SNe detonate over a range of delay times sampled from {\sc Starburst99} \citep{Leitherer2014}. In order to prevent the overcooling \citet{Kimm2015} found for long ($10~\Myr$) SNe delays, we choose a smaller timescale, comparable to the feedback disruption times observed in \cite{Kruijssen2019} and \cite{Chevance2020}. Each SNe also ejects $10.5~\Msun$ of mass ($M_{\rm ej}$) and $2~\Msun$ of metal ejecta, the same total amounts used in the {\sc FIRE-2} simulations \citep{Hopkins2018b}.

\subsection{Initial Conditions} \label{ss:ICs}

The initial conditions of our simulations have been created following the procedure outlined in \cite{Springel2005}. Each isolated galaxy consists of a stellar, dark matter, and gas component, with the stellar component possibly subdivided into a disc and a bulge component. All initial conditions are based on the standard \textsc{Agora} disc \citep{Kim2016}. However, to assess how different gravitational potentials influence the SFR and whether bulges can suppress star formation, a set of initial conditions with a variety of bulge mass fractions and scale radii is needed. Thus, while otherwise similar to the initial conditions used in the \textsc{Agora} disc, the bulge component can differ. Below, the components defining our fiducial model are described in detail. Table~\ref{tab:ICs} lists the full suite of simulations and the quantities that we vary across the suite.

The dark matter halo is modelled as a \cite{Hernquist1990} profile, with a concentration parameter of $c = 10$, a spin parameter of $\lambda = 0.04$, and a circular velocity $v_{\rm circ} = 180~\kms$, similar to those of the Milky Way \citep{Bland-Hawthorn2016}. A \cite{Hernquist1990} profile is also used to describe the bulge component. It is defined by the bulge mass $\Mb$ and its scale radius $\Rb$. To explore the effect of varying the gravitational potential, we vary both of these parameters. In order to ensure that all changes are due to a change in the bulge component, all galaxies in the sample have the same initial total stellar mass, of $\Mstar \sim 4.71 \times 10^{10}~\Msun$. We split this mass into a bulge and disc component, i.e.\ $\Mstar = \Mb + \Md$, and include a control run that only has a disc ($\Mb=0$). We consider bulge mass fractions of 30, 60, and 90~per~cent of the initial stellar mass, and scale radii of 1, 2, and 3~kpc. Our fiducial model is chosen to have a bulge component with a mass and radius in the middle of the parameter space covered in this exploration, i.e.\ the scale radius is $\Rb = 2\kpc$ and the bulge contains 60 per cent of the initial stellar mass, yielding a $\Mb/\Md$ ratio of 1.5. The stellar disc is described by an exponential radial profile with an initial scale length $R_{\rm d} = 4.6~\kpc$, and a vertical $\rm sech^2$ profile with scale height of $0.1 R_{\rm d}$.

We express the initial amount of gas as ratio of gas to the total stellar mass, because the disc mass varies greatly between the initial conditions. For simplicity (to not include another dimension into the parameter space) we choose the same gas fraction for all galaxies in this study. To mimic the relative gas-poorness of early-type galaxies \citep{Young2011} and in agreement with the findings of \cite{Saintonge2017} for galaxies of similar stellar mass, the initial total gas to stellar mass ratio is fixed to $\Mgas/\Mstar=0.05$.

\begin{table*}
 \begin{tabular}{lcccccc}
  \hline
  Name & M$_{\rm b}$ [10$^{10}\Msun$] & R$_{\rm b}$ [kpc] & Resolution [$\Msun$] &  SFE model & Remark \\
  \hline
  noB    &  0 &  0 & $1 \times 10^{4}$ &  $\eff = \rm f(\avir)$ & `bulgeless'\\
  B\_M30\_R1 & 1.41 & 1 & $1 \times 10^{4}$ &  $\eff = \rm f(\avir)$ & \\
  B\_M30\_R2 & 1.41 & 2 & $1 \times 10^{4}$ &  $\eff = \rm f(\avir)$ & \\
  B\_M30\_R3 & 1.41 & 3 & $1 \times 10^{4}$ &  $\eff = \rm f(\avir)$ & \\  
  B\_M60\_R1 & 2.83 & 1 & $1 \times 10^{4}$ &  $\eff = \rm f(\avir)$ & \\
  B\_M60\_R2 & 2.83 & 2 & $1 \times 10^{4}$ &  $\eff = \rm f(\avir)$ & `fiducial' \\
  B\_M60\_R3 & 2.83 & 3 & $1 \times 10^{4}$ &  $\eff = \rm f(\avir)$ & \\
  B\_M90\_R1 & 4.24 & 1 & $1 \times 10^{4}$ &  $\eff = \rm f(\avir)$ & `compact bulge' \\
  B\_M90\_R2 & 4.24 & 2 & $1 \times 10^{4}$ &  $\eff = \rm f(\avir)$ & \\
  B\_M90\_R3 & 4.24 & 3 & $1 \times 10^{4}$ &  $\eff = \rm f(\avir)$ & \\  
  noB\_cSFE &  0 &  0 & $1 \times 10^{4}$ &  $\eff = 1$~per~cent &  \\
  B\_M60\_R2\_cSFE & 2.83 & 2 & $1 \times 10^{4}$ &  $\eff = 1$~per~cent &  \\
  B\_M90\_R1\_cSFE & 4.24 & 1 & $1 \times 10^{4}$ &  $\eff = 1$~per~cent &  \\
  \hline
  B\_M60\_R2\_hres & 2.83 & 2 & $3 \times 10^{3}$ &  $\eff = \rm f(\avir)$ & `high resolution'\\
  B\_M60\_R2\_lres & 2.83 & 2 & $3 \times 10^{4}$ &  $\eff = \rm f(\avir)$ & `low resolution'\\
  \hline
  \hline
 \end{tabular}
 \caption{Initial conditions of the simulations. 
 The naming convention of the simulations is to first list the presence of a bulge (`B' or `noB'), followed by the relative bulge mass (`M$X$' with $X$ the percentage of the total mass constituted by the bulge) and then the bulge scale radius (`R$Y$' with $Y$ the radius in kpc). Runs with a constant $\eff$ have `cSFE' appended; similarly, the postfix `res' indicates runs in which the resolution is varied. The final column lists the descriptive designations by which we refer to some of the simulations throughout this paper.}
 \label{tab:ICs}
\end{table*}

With an average mass resolution of $\sim 1 \times~10^{4}\Msun$ and a density threshold for star formation of $1~\ccm$, we use a minimum gravitational softening length of 12~pc. With these choices, we ensure that gas cells are gravitationally resolved for densities up to two orders of magnitude higher than the density threshold. This is relevant in the context of our star formation model, because it ensures that the gas can condense into structures with densities higher than the threshold before becoming self-gravitating. The gravitational softening for the dark matter halo is coarser, with a softening length of 26~pc. The gravitational softening is modified for the resolution tests, with minimum softening lengths of 15~(35) and 6~(12)~pc for baryons (dark matter) in the runs B\_M60\_R2\_lres and B\_M60\_R2\_hres, respectively.  

\begin{figure*}
    \centering
    \includegraphics[width=1.\linewidth]{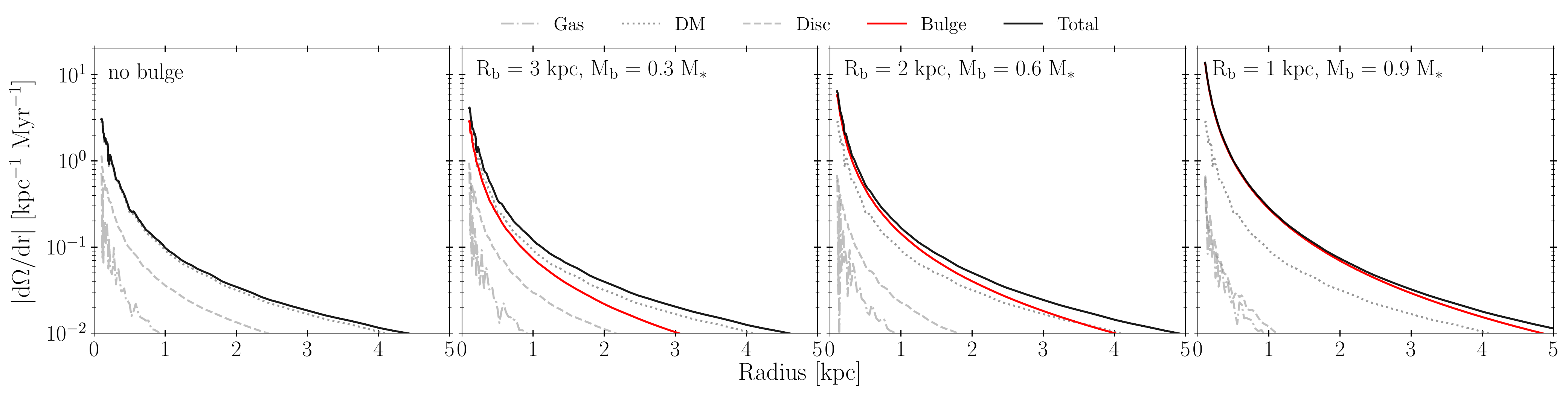}
    \includegraphics[trim=5mm 0mm 1.5mm -5mm, clip=true,width=1.\linewidth]{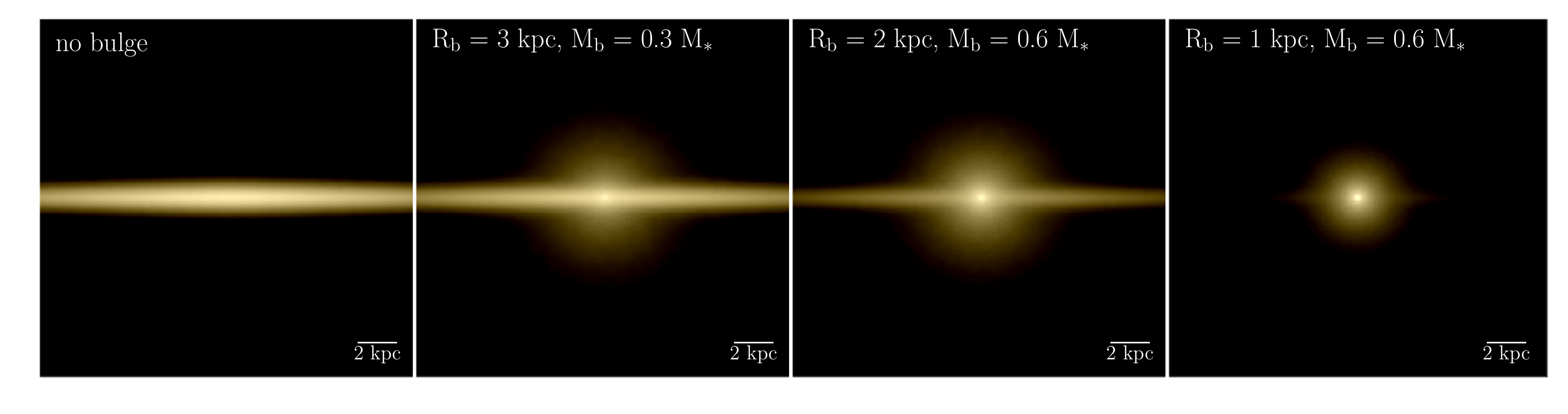}%
    \caption{Gradient of the angular velocity $|{\rm d}\Omega/{\rm d}r|$ (a measure of shear) as a function of galactocentric radius (top) and mock colour images (using $u$, $v$, $i$ filters) of the stars (bottom) for four initial conditions from our simulation suite. From left to right, these show the bulgeless model (noB), the weakest bulge model (B\_M30\_R3), the fiducial bulge model (B\_M60\_R2), and most dominant bulge model (B\_M90\_R1). For the first two of these models, the dark matter halo is the dominant source of shear. The total shear increases in the presence of a bulge; the more prominent the bulge component, the stronger the effect. The shear experienced by the fiducial run is dominated by the bulge component in the inner $3.5~\kpc$. For the most compact bulge, the shear induced by the bulge component dominates over the contribution from the dark matter out to 7~kpc.   
    }
      \label{fig:ex_vcirc_fcimg}
\end{figure*}

We show a selection of initial conditions in Figure~\ref{fig:ex_vcirc_fcimg}, to give the reader an idea of how the stellar component looks visually, as well as to quantify the impact of the different bulge radii and bulge/disc ratios on the shear experienced by the ISM in these galaxies. More massive and more compact (i.e.\ higher density) bulges increase the shear, most strongly so in the centre, but also throughout the galaxy. For the same reason, the galactocentric radius out to which the bulge dominates the galactic shear is larger for higher bulge densities.

\section{Comparison of star formation models with a constant or varying SFE} \label{s:cvSFE}

\subsection{Star Formation} \label{ss:vcSFE}

\begin{figure*}
    \centering
    \includegraphics[width=1.\linewidth]{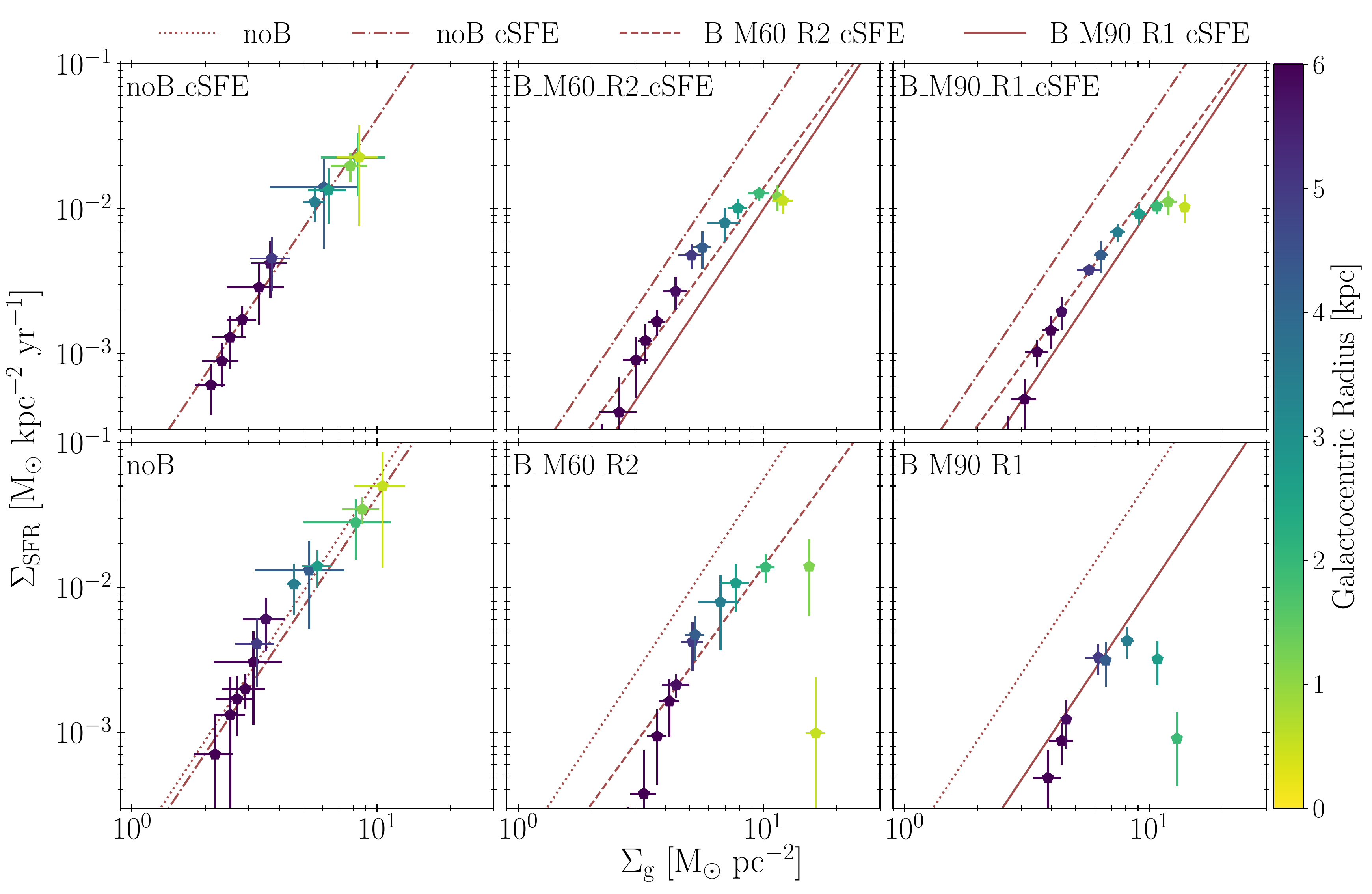}%
    \caption{
    SFR surface density as a function of gas surface density for the bulgeless (left), fiducial (centre) and compact-bulge (right) simulations, contrasting the constant (top) and dynamics-dependent (bottom) SFE models. The colour coding indicates the galactocentric radius. Red lines show the best-fitting power law relations for the panels indicated in the legend. Each point represents a time average of snapshots separated by $100~\Myr$; the snapshot-to-snapshot variation is shown by the error bars. The central $300~\pc$ are excluded from the analysis (see Section~\ref{ss:vcSFE}). This figure shows that the presence of a spheroid suppresses the SFR towards small galactocentric radii in all simulations, but most strongly so for the dynamics-dependent star formation model.} 
    \label{fig:vcSFE_KS}
\end{figure*}

To assess how the global SFR is affected by the dynamics-dependent and constant SFE models, as well as by different stellar potentials, we consider the radially-binned star formation relation of a subset of six simulations, shown in Figure~\ref{fig:vcSFE_KS}. Following \cite{Kruijssen2014a}, who show that the temporal and spatial variations of the star formation relation caused by cloud evolution introduce considerable scatter on sub-kpc scales, we calculate the SFR and gas surface density in radial annuli of 750~$\pc$ width. 
We exclude the central $300~\pc$ from our analysis, because we omit feedback from active galactic nuclei and lack the resolution to accurately model the star formation activity in the very nucleus. Though mostly relevant for the spheroid-dominated galaxies, we apply this cut to all galaxies in the sample. We refer to Section~\ref{ss:vSFE_avir} and \ref{ss:numcavs} for more details. In analogy to observations \citep[e.g.][]{Kennicutt2012,Leroy2013,Haydon2018}, we only use the stars which formed in the past 10~$\Myr$ to calculate the mean SFR over this time interval. To obtain a measure of the variation in $\Ssfr$ and $\Sg$ we average over snapshots separated by 100~$\Myr$; this variation is indicated by the error bars in Figure~\ref{fig:vcSFE_KS}. We remind the reader that all simulated galaxies have a gas fraction of 5~per~cent.

Both bulgeless runs follow a similar star formation relation and are largely insensitive to the sub-grid star formation model. The net SFR of the dynamics-dependent efficiency run is slightly higher, as indicated by the different normalisation. However, within the error bars arising from time variation, the SFRs of the simulations agree remarkably well with each other. A similar star formation relation is seen for the B\_M60\_R2\_cSFE simulation, but its overall SFR is lower, indicated by the offset between the best fit lines in the upper middle panel of Figure~\ref{fig:vcSFE_KS}. This reflects the stabilising effect of the bulge (see below).
The dynamics-dependent $\eff$ model in the bottom middle panel shows a similar relation down to the inner $1.5~\kpc$. At smaller radii, the B\_M60\_R2 run shows a pronounced drop in SFR, similar to those observed by \cite{MendezAbreu2019}. Differences in SFR resulting from different sub-grid models are even starker for the B\_M90\_R1 and B\_M90\_R1\_cSFE simulations. Run B\_M90\_R1\_cSFE effectively follows the same star formation relation as run B\_M60\_R2\_cSFE. However, the SFR of run B\_M90\_R1 peaks at a radius of $\sim 3.5~\kpc$ before decreasing again, despite increasing gas surface densities, showing a much stronger suppression of star formation than in the fiducial run or the compact bulge run with a constant SFE.

The above comparison highlights why models with a constant SFE have been so widely used. For a disc galaxy without a pronounced spheroid, the SFR is similar to that obtained with a dynamics-dependent star formation model, irrespective of the choice of $\eff$, density threshold, and temperature ceiling, as expected from the self-regulation of star formation and feedback \citep{Agertz2013}. However, in order to successfully reproduce the suppressed SFRs towards the centres of bulge-dominated galaxies \citep[e.g.][]{Longmore2013,Davis2014}, a dynamics-dependent sub-grid model is required.

We expect that the implications of changing the sub-grid star formation model extend beyond the star formation relation, because the feedback resulting from star formation will impact the gas differently depending on where, when and at which rate the stars are formed. In turn, this will impact the ISM structure and kinematics, thereby influencing future star formation. Therefore, we now proceed with a detailed analysis of how the cloud-scale baryon cycle of ISM evolution, star formation, and feedback is affected by the star formation model and galactic morphology.

\begin{figure*}
    \centering
    \includegraphics[width=1.\linewidth]{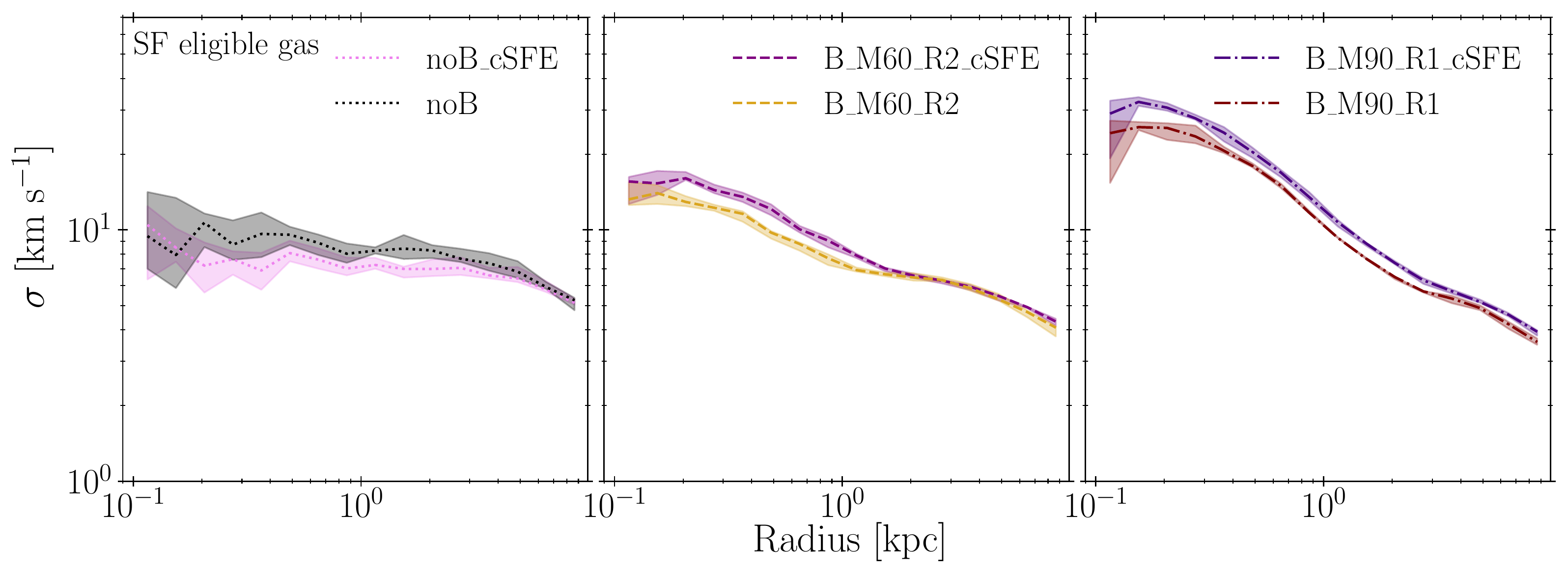}%
    \caption{
    Comparing the effect of different sub-grid star formation models on the turbulent velocity dispersion of the gas, for the bulgeless (left), fiducial (middle) and compact bulge (right) galaxies. The panels show the result for gas that satisfies the density and temperature thresholds for star formation.
    }
    \label{fig:vcSFE_veldisp_radprof}
\end{figure*}
    
\subsection{Effect on ISM properties} \label{ss:vcSFE_GPE}
We now turn to an analysis of the impact of different star formation models on the state of the turbulent ISM. Specifically, we consider the gas velocity dispersion and the virial parameter, which directly enter into the dynamics-dependent star formation model. In addition, we consider the turbulent pressure, because it is a crucial component for balancing self-gravity and maintaining hydrostatic equilibrium. We also consider the ISM morphologies and gas surface density profiles of the modelled galaxies, because these set the SFR and are likely to be affected by changes in the star formation model. 

To ensure a representative and reproducible analysis of these properties, we measure them in different snapshots and use the median and 16th-to-84th percentiles over these snapshots. This allows us to quantify the overall trends, as well as to quantify the stochastic variation in time introduced by the quantisation of star formation and feedback into individual events \citep[e.g.][]{Kruijssen2018}. We use snapshots starting at $300~\Myr$, to allow the galaxies to settle into equilibrium, and subsequent ones separated by $100~\Myr$ (roughly a galactic dynamical time), to make sure that snapshots are independent. We run the simulations for a Gyr, resulting in a total of eight snapshots combined this way. Unless explicitly stated otherwise, we will use this approach to calculate any quantities throughout the rest of the paper. 

\subsubsection{Velocity dispersion} \label{sss:vcSFE_veldisp}

We show the radial gas velocity dispersion profiles in Figure~\ref{fig:vcSFE_veldisp_radprof}, where we compare simulations run with sub-grid models using a constant and dynamics-dependent efficiency, but with the same (stellar) gravitational potential. The velocity dispersion in galaxy discs is set by a combination of the gravitational potential and stellar feedback \citep{Krumholz2018}. Because we fix the potential, differences between models can only be the consequence of differences between the star formation models. While the star formation prescription itself cannot directly affect the velocity dispersion, differences in the resulting SFRs imply the injection of different amounts of total energy and momentum by stellar feedback. This means that the star formation model indirectly changes the gas velocity dispersion through stellar feedback.

\begin{figure*}
    \centering
    \includegraphics[width=1.\linewidth]{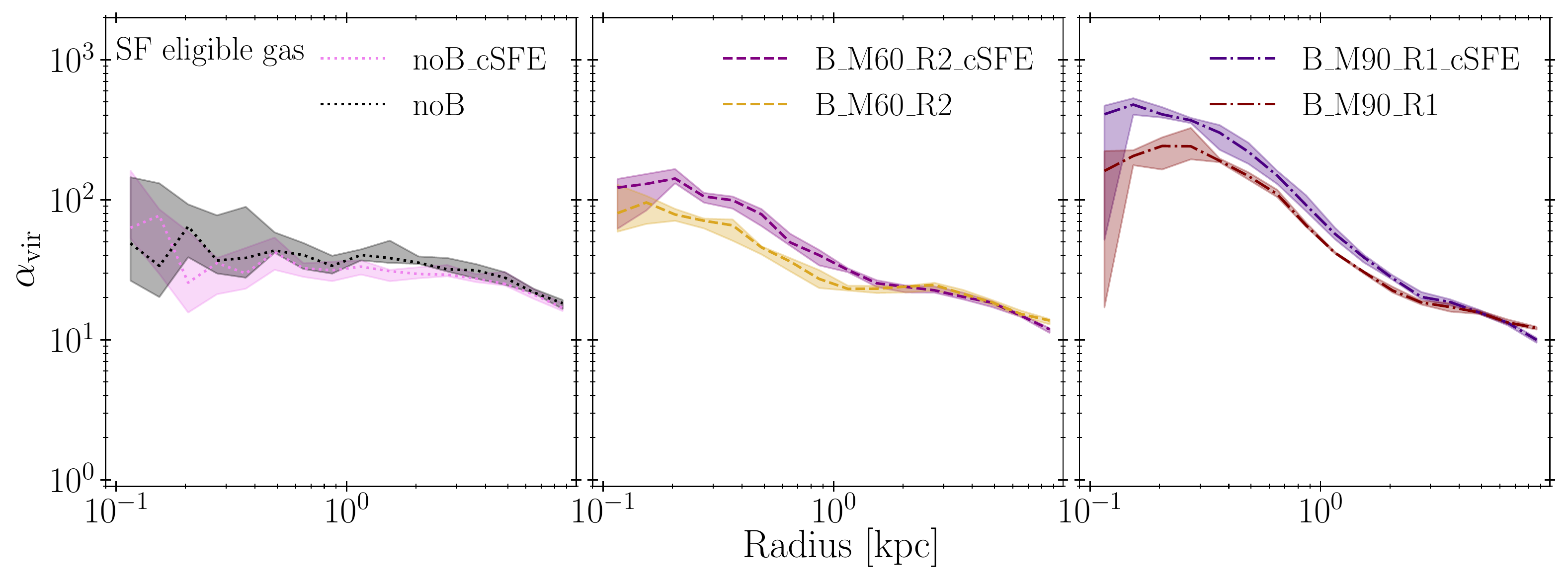}%
    \caption{
    Comparing the effect of different sub-grid star formation models on the virial parameter of the gas, for the bulgeless (left), fiducial (middle) and compact bulge (right) galaxies. The panels show the result for gas that satisfies the density and temperature thresholds for star formation. 
    }
    \label{fig:vcSFE_avir_radprof}
\end{figure*}

Considering the bulgeless simulations first, the velocity dispersions remain approximately flat in the inner $\sim5~\kpc$ of the galaxy. This is seen more strongly for the simulations with a constant SFE. The dynamics-dependent efficiency run has a higher absolute SFR, causing the gas in this galaxy to be more turbulent. However, these differences remain largely within the variation of the median $\sigma$ over time. The variance of the median does increase towards the centre of the galaxy for both star formation models. This increase is driven by the greater degree of stochasticity towards the galactic centre, where the star formation is burstier.

Focussing on the differences between sub-grid star formation models in the presence of a bulge (middle and right panels of Figure~\ref{fig:vcSFE_veldisp_radprof}), three points become apparent. Firstly, irrespectively of the sub-grid star formation model and the (subset of) gas considered, the velocity dispersion increases towards the centre of the galaxy, specifically within the inner $1{-}3$~kpc. This is a result of the deeper central stellar potential in bulge-dominated galaxies. For a more elaborate exploration of how the underlying gravitational potential affects $\sigma$, we refer to Section~\ref{ss:vSFE_veldisp}. Secondly, there is a small but distinct difference in $\sigma$ between the different sub-grid models, within the respective bulge-dominated regions. These are the inner $1{-}2~\kpc$ for the fiducial models, and nearly all of the galaxy (but most prominently the inner $3{-}4~\kpc$) for the most dominant bulge. In the simulations with a constant SFE, the SFR is not (as) strongly suppressed in the central regions. The resulting, higher momentum and energy input from stellar feedback is responsible for the offset in velocity dispersions. It adds to the increase of the velocity dispersion caused by the presence of a bulge, and causes the central rise of the velocity dispersion to extend further out into the disc. Thirdly, the variance of the velocity dispersion is smaller in galaxies with a bulge than in the simulations without a bulge. This relates back to the star formation within the galaxy. Those galaxies with a high(er) SFR do so because they experience larger extremes, with more subsequent feedback events, leading to a larger variation in $\sigma$ over time.

In conjunction with Figure~\ref{fig:vcSFE_KS}, Figure~\ref{fig:vcSFE_veldisp_radprof} highlights the necessity of a dynamics-dependent star formation model. The SFRs of runs B\_M90\_R1\_cSFE and B\_M60\_R2\_cSFE are nearly the same, despite very different gas velocity dispersions. Only B\_M60\_R2 and B\_M90\_R1 reproduce the trend obtained from cloud-scale simulations that highly turbulent, super-virial gas should form stars less efficiently \citep[e.g.][]{Federrath2012,Padoan2012,Padoan2017}.

\subsubsection{Virial Parameter} \label{sss:vcSFE_avir}

Next, we compare the effect of the sub-grid star formation model on the virial parameter. As $\avir \propto\sigma^2$, the virial parameter is expected to show similar trends with radius as the velocity dispersion. This is indeed seen in Figure~\ref{fig:vcSFE_avir_radprof}. 

\begin{figure*}
    \centering
    \includegraphics[width=1.\linewidth]{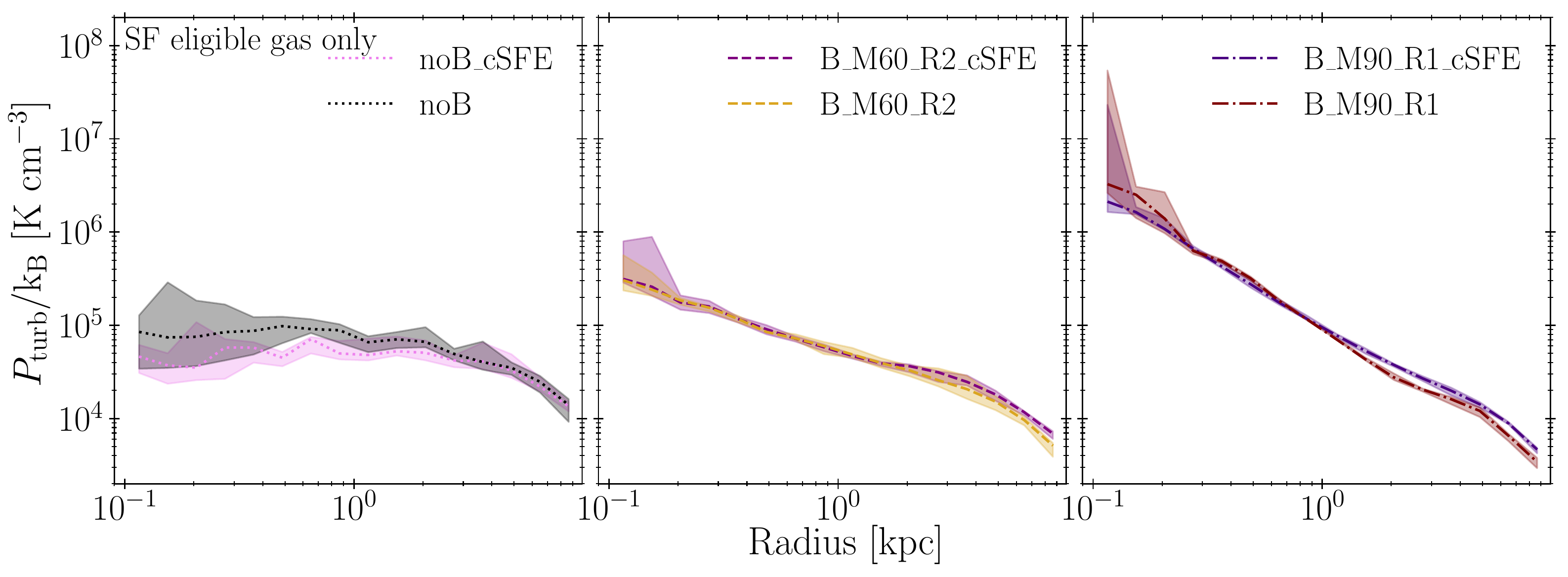}%
    \caption{
    Comparing the effect of different sub-grid star formation models on the turbulent pressure of the gas, for the bulgeless (left), fiducial (middle) and compact bulge (right) galaxies. The panels show the result for gas that satisfies the density and temperature thresholds for star formation.
    }
    \label{fig:vcSFE_pturb_radprof}
\end{figure*}

When considering the gas eligible for star formation, the gas virial parameters in the bulgeless galaxies increase a little from the disc to the centre. The simulations with a constant SFE show a more pronounced upturn in $\avir$ within the inner 200 $\pc$, but these differences fall largely within the temporal variance of the models. The median virial parameters are slightly higher in the simulations with a dynamics-dependent $\eff$, which mirrors the slight elevation of the velocity dispersion caused by the higher SFR and the resulting stellar feedback. 

For the simulations with the fiducial bulge, the difference between the virial parameters predicted by the two sub-grid star formation models near the galaxy centres is larger than that between the velocity dispersions in Figure~\ref{fig:vcSFE_veldisp_radprof}, even after accounting for the fact that $\avir\propto\sigma^2$. This is related to an additional dependence of $\avir\propto L/M$, which means that a higher gas density lead to a lower virial parameter at fixed cloud size. While the gas in the centre of the galaxy is less turbulent for the dynamics-dependent model, it is also more dense, due to the prominent absence of star formation, and the fact that clouds near galaxy centres are more compact due to the elevated tidal field strength, shear, and geometric convergence \citep[e.g.][]{Kruijssen2019b}. All of these factors result in the median virial parameter being lower in simulations run with a dynamics-dependent SFE. 

The trends for the most bulge-dominated galaxies are similar to those of the fiducial runs. Together, all panels sketch a picture in which the presence of a bulge elevates the gas virial parameters towards the galactic centre. This increase is slightly less pronounced when using a dynamics-dependent SFE, due to the suppression of star formation and any subsequent feedback by the elevated virial parameter (see equation~\ref{eq:eff}).

\subsubsection{Turbulent Pressure} \label{sss:vcSFE_Pturb}

In Figure~\ref{fig:vcSFE_pturb_radprof}, we show how the turbulent pressure, $\Pt = \rho \sigma^2$, of the gas changes between the different sub-grid models for star formation. In the bulgeless simulations, the turbulent pressure increases inwards throughout the disc, until it flattens within the inner $2~\kpc$. Due to the overall similar SFR (and as with $\sigma$ and $\avir$), the profiles of the different sub-grid star formation models agree within the variation of the median over time. The median $\Pt$ of gas in the fiducial and compact bulge models keeps increasing towards the centre without such flattening. This is caused by the additional hydrostatic pressure generated by the bulge, which the turbulent pressure equilibriates to \citep[e.g.][]{Schruba2019}.

\subsubsection{ISM morphology and gas surface density profile}

\begin{figure*}
    \centering
    \includegraphics[trim=48mm 30mm 6mm 25mm, clip=true, width=1.\linewidth]{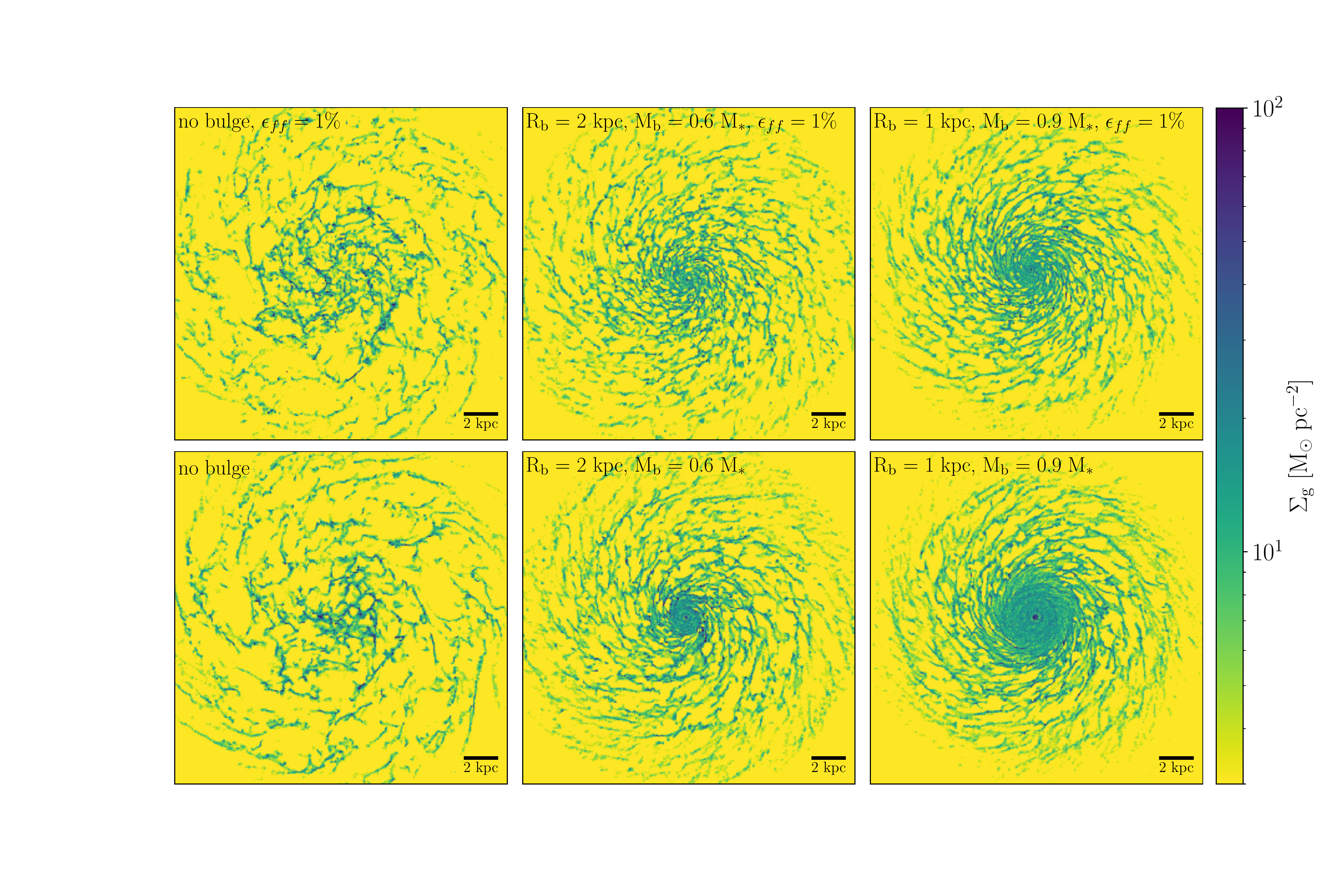}%
    \caption{
    Surface density projection of the gas discs for the bulgeless (left), fiducial (middle) and compact bulge (right) galaxies. The top panels show the result for galaxies simulated with a constant SFE sub-grid star formation model, whereas the bottom panels show the dynamics-dependent analogue. The maps are shown at 600~Myr after the start of the simulations and measure $20~\kpc$ on a side. The unsuppressed SFR and subsequent feedback in both bulgeless simulations leads to a flocculent ISM, with most of the gas along spiral arms, whereas the bulge-dominated galaxies host smooth, quiescent central gas discs. 
    }
    \label{fig:vcSFE_sdproj}
\end{figure*}

\begin{figure*}
    \centering
    \includegraphics[width=1.\linewidth]{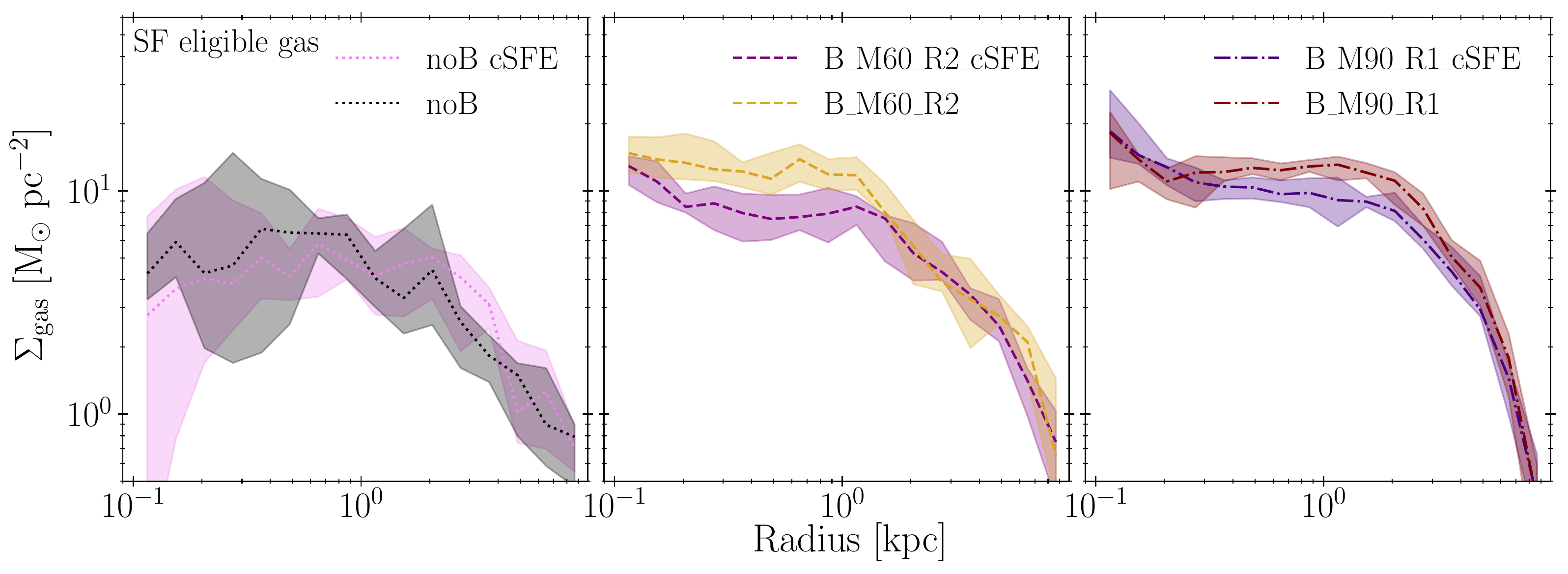}%
    \caption{
    Comparing the effect of different sub-grid star formation models on the gas surface density profile, for the bulgeless (left), fiducial (middle) and compact bulge (right) galaxies. The panels show the result for gas that satisfies the density and temperature thresholds for star formation.
     }
    \label{fig:vcSFE_sd_radprof}
\end{figure*}

We show the effect of the star formation model on the distribution of gas within the galaxy in Figure~\ref{fig:vcSFE_sdproj}. The bulgeless galaxies look very similar in projection. The molecular ($\Sg\ga10~\Msun~\pc^{-2}$) gas is distributed along thin arm-like structures and the centre exhibits considerable sub-structure. Thus, while undergoing a lot of variation, the median surface density as a function of radius, shown in Figure~\ref{fig:vcSFE_sd_radprof}, is similar at all radii.

Both fiducial bulge galaxies host a larger molecular gas reservoir in their centre. However, the central $\sim 2\kpc$ of the constant $\eff$ galaxy are more sub-structured than for the galaxy with the dynamics-dependent $\eff$. Because galaxy B\_M60\_R2\_cSFE forms (more) stars than B\_M60\_R2 at all radii, including in the centre, stellar feedback is more effective at disrupting the gas. In the dynamics-dependent model, where feedback is weaker due to the low SFR, the gas is able to settle into a smooth, high-density disc at the centre. The compact bulge simulations behave very similarly. The gas disc in run B\_M90\_R1 is even larger and smoother than that of B\_M60\_R2. Run B\_M90\_R1\_cSFE has a small disc at the very centre, but at a lower density and with more substructure than in B\_M90\_R1.

The behaviour seen in the gas surface density maps is reflected in the radial gas surface density profiles in Figure~\ref{fig:vcSFE_sd_radprof}, where the median $\Sg$ of B\_M60\_R2 and B\_M90\_R1 is consistently offset to larger surface densities with respect to the analogous simulations with a constant SFE. Again, this difference results from the differing SFRs and corresponding feedback intensities introduced by taking galactic dynamics into account.

In principle, it could be argued that the specific value of $\eff$ in the constant SFE model could be adjusted to yield SFRs and galaxy properties more similar to those found for a dynamics-dependent star formation model. Alternatively, the overall offsets may be considered to be too small given the numerical and observational uncertainties. However, taken at face value, the differences shown in this section highlight the importance of carefully choosing the sub-grid star formation model and associated parameters, as the consequences reach beyond global scalings of the SFR and instead affect the structural evolution of the galaxy and its ISM at large. We furthermore stress that significantly suppressed star formation in a bulge-dominated galaxy is only reproduced when explicitly accounting for the effect of galactic dynamics in the sub-grid star formation model, which then in turn introduces further differences. Conversely, galaxies evolved with the constant SFE model do not reflect any of the changes in ISM kinematics that result from the presence of different types of bulge.

\section{Effect of the gravitational potential on ISM properties} \label{s:BC}

We now proceed to only consider the dynamics-dependent star formation model and vary the gravitational potential across the comprehensive parameter space covered by our simulations. We then compare how the variety of ISM-related quantities from Section~\ref{s:cvSFE} differ between these simulations. This comparison is aimed at revealing the physical mechanisms that link galactic morphology to ISM properties, star formation, and quenching, with the eventual goal of quantifying how these effects may change the star formation relation in these galaxies. In Section~\ref{s:Dis}, we will demonstrate how the dynamical suppression of star formation can drive entire galaxies off the main sequence of star-forming galaxies, into the red cloud of quenched systems.

\subsection{Turbulent velocity dispersion} \label{ss:vSFE_veldisp}

\begin{figure*}
    \centering
    \includegraphics[width=1\linewidth]{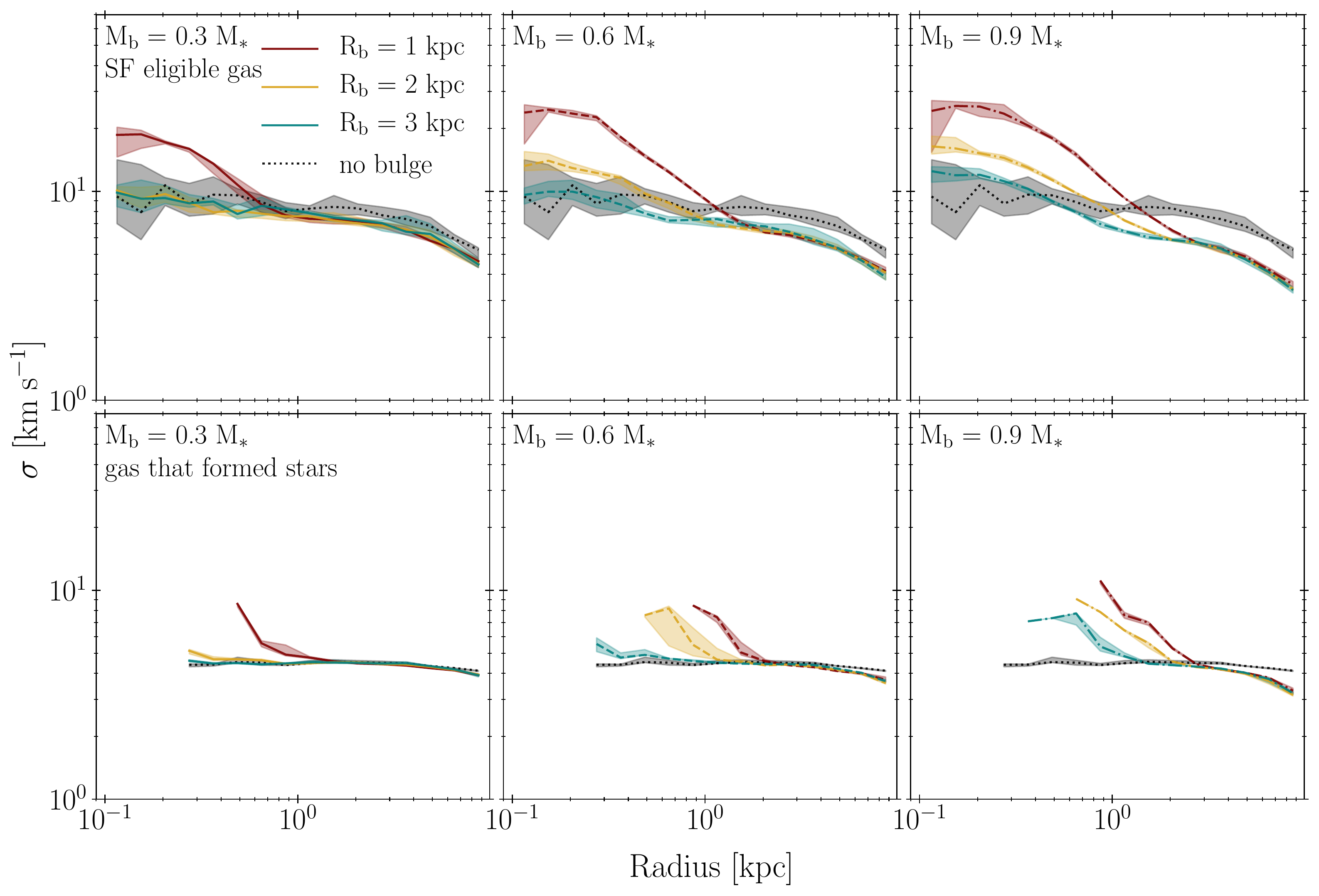}%
    \caption{
    Radial profiles of the gas turbulent velocity dispersion for different gravitational potentials. Each panel compares the effect of different bulge scale radii at a constant bulge mass, with the columns showing simulations with bulges containing 30, 60 and 90~per~cent of the initial stellar mass, from left to right. The top row shows the result for gas that satisfies the density and temperature thresholds for star formation, whereas the bottom row only includes gas from which stars formed. The central $300~\pc$ are excluded from the analysis of the bottom panels (see Section~\ref{ss:vcSFE}). Each panel includes a dotted line showing the bulgeless simulation. 
    } 
    \label{fig:veldispprof_vSFE}
\end{figure*}

The velocity dispersion of the gas is one of the main gas properties setting the virial parameter (and thus the SFE). In Figure~\ref{fig:veldispprof_vSFE}, we show the radial profiles of $\sigma$ for all simulations with a dynamics-dependent SFE. The bulgeless simulation is included in every panel as a reference line. Because the bulgeless galaxy is a pure exponential disc, the median velocity dispersion is approximately flat as a function of radius, only declining slightly in the outskirts of the disc.

From the velocity dispersion profiles displayed in Figure~\ref{fig:veldispprof_vSFE} it is clear that both $\Rb$ and $\Mb$ have an influence on the dynamical state of the gas.
All three mass fractions exhibit a similar behaviour, in that the velocity dispersion of the gas increases towards the centre of the galaxy in the presence of a bulge. The more centrally concentrated the bulge (i.e.\ the smaller $\Rb$ is), the stronger the effect. Bulges with smaller $\Rb$ induce more turbulence at a given $\Mb$. Additionally, gas further away from the centre is affected by the potential. It is likely that the true dependence on bulge properties uses a combination of $\Rb$ and $\Mb$, such as through a linear, surface, of volume mass density enclosed at each radius (reflecting the potential, force, and tidal field, respectively). In either of these cases, we expect a monotonic trend of ISM properties and SFR with $\Mb/\Rb^n$ with $n=1{-}3$. We demonstrate such a dependence of the SFR on the enclosed mass surface density ($n=2$) in Section~\ref{ss:obscomp}.

The set of galaxies with the lowest bulge fraction, $\Mb = 0.3\Mstar$ shows the smallest differences in velocity dispersion. Only the $\Rb=1~\kpc$ bulge displays a strong $\sigma$ increase in the inner few hundred pc, by $\la 0.3$~dex. At this bulge mass, the two more extended bulges have very similar velocity dispersion profiles, largely resembling the bulgeless galaxy. Based on the discussion so far, this may seem surprising, because one would expect the B\_M30\_R2 run to exhibit a larger velocity dispersion than the B\_M30\_R3 run. However, as can be seen in the rotation curves of Figure~\ref{fig:ex_vcirc_fcimg}, the galaxy with the weakest bulge resembles the bulgeless run very closely. It has the second largest net SFR of all simulations with a dynamics-dependent $\eff$. The resultant feedback is sufficient to increase $\sigma$ to same level as in run B\_M30\_R2.
At bulge mass fractions of 60 and 90~per~cent, the effect of the bulge scale radius becomes more discernible, because the bulge potential is no longer drowned out by the dark matter halo. Even run B\_M60\_R3 shows a higher velocity dispersion towards the centre than in its outskirts, although the total dynamic range of $\sigma$ is again only of order $\sim 0.3$~dex. For the most dominant bulge (B\_M90\_R1), the effect of the potential on $\sigma$ is noticeable out to $3~\kpc$, increasing the velocity dispersion by nearly an order of magnitude when going from the outer disc to the centre of the galaxy.

Qualitatively, the above trends hold irrespectively of whether we consider all gas eligible for star formation, or only that from which stars have formed. Because we exclude the central $300~\pc$ from the star formation analysis, we exclude them in the profiles of gas that formed stars, too. The only difference between both subsets of gas cells is that the median velocity dispersions of the star-forming gas are $\sim 0.4$~dex lower than those of all star formation eligible gas, since star formation preferentially proceeds in regions with lower velocity dispersions \citep[e.g.][]{Padoan2012}. This means that, irrespectively of the gas tracer used, both the increase of the gas velocity dispersion towards small galactocentric radii and the overall larger values of $\sigma$ for the most massive, centrally concentrated spheroids are strong, observationally testable predictions of our simulations. These trends are key empirical diagnostics signposting a strong interplay between galactic dynamics and ISM kinematics in galactic spheroids and early-type galaxies. 

\subsection{Virial Parameter} \label{ss:vSFE_avir}

\begin{figure*}
    \centering
    \includegraphics[width=1.\linewidth]{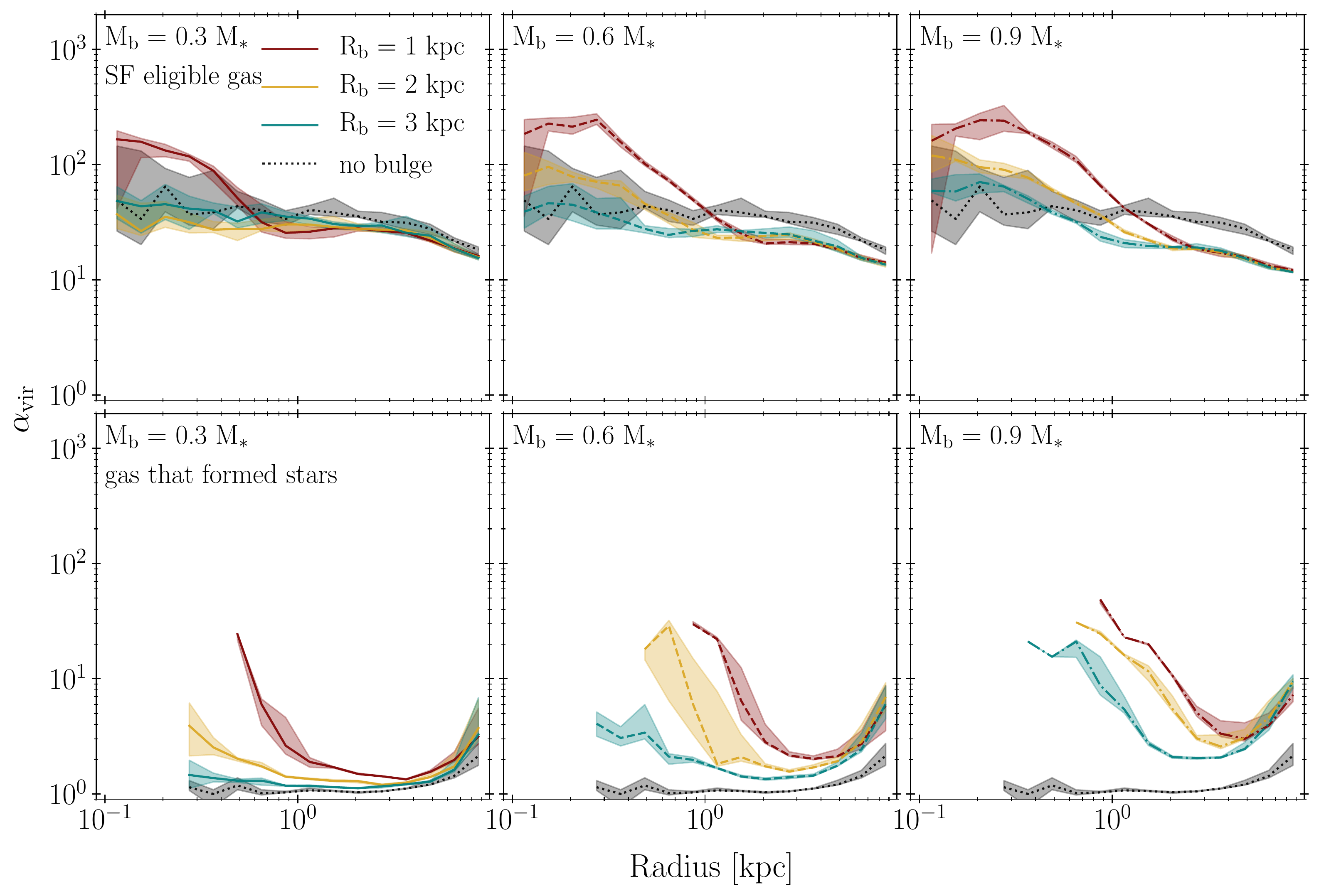}%
    \caption{
    Radial profiles of the gas virial parameter for different gravitational potentials. Each panel compares the effect of different bulge scale radii at a constant bulge mass, with the columns showing simulations with bulges containing 30, 60 and 90~per~cent of the initial stellar mass, from left to right. As before, the top row shows the result for gas that satisfies the density and temperature thresholds for star formation, whereas the bottom row only includes gas from which stars formed. The central $300~\pc$ are excluded from the analysis of the bottom panels (see Section~\ref{ss:vcSFE}). Each panel includes a dotted line showing the bulgeless simulation.
    } 
    \label{fig:avirradprof_vSFE}
\end{figure*}

Figure~\ref{fig:avirradprof_vSFE} shows that the virial parameter follows the same trends as the velocity dispersion with respect to bulge mass and scale radius. Resulting from the increase in turbulent velocity dispersion towards the centres of the galaxies with a bulge, $\avir$ also increases towards the centres, and more strongly so for more compact and massive bulges. Quantitatively, the difference in $\avir$ between the centre and the outskirts of the disc in a bulge-dominated galaxy can reach around an order of magnitude for the most compact bulges. For the most compact bulge, the  difference remains up to $\sim 0.3$~dex when contrasting the virial parameter at radii of $2.5{-}3~\kpc$ to the outer disc. Again, the bulgeless galaxy has a much flatter profile, with a higher median $\avir$ in the outer disc.

\begin{figure*}
    \centering
    \includegraphics[trim=10mm 0mm 30mm -0.1mm, clip=true, width=1.\linewidth]{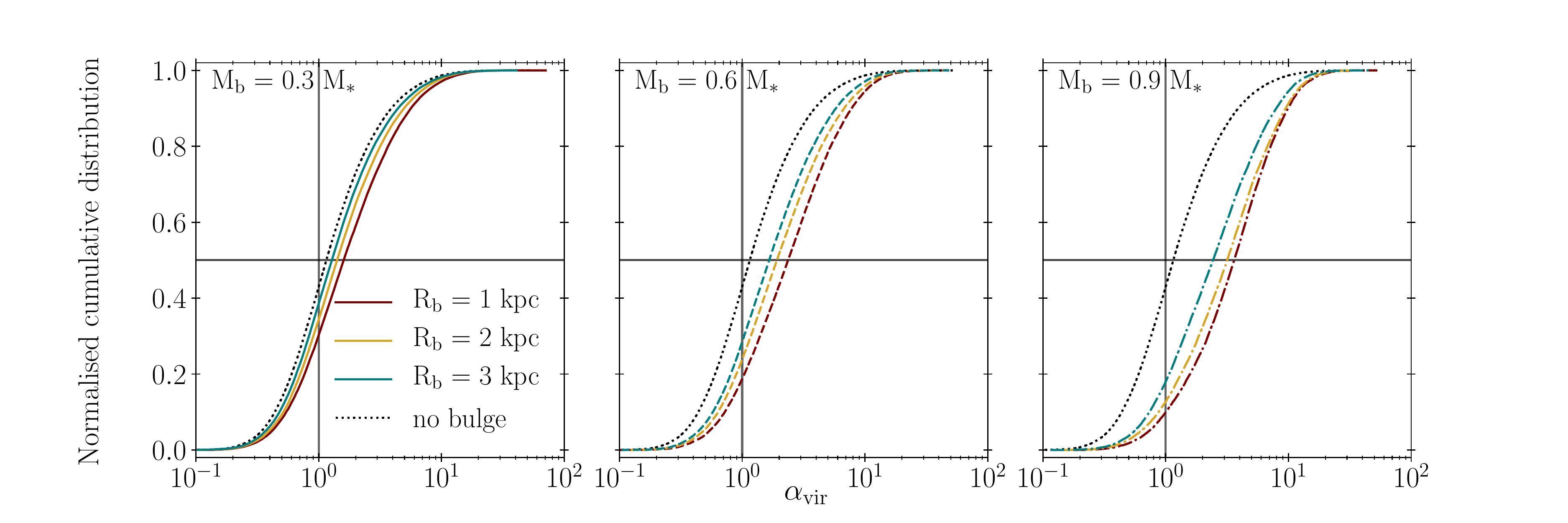}%
    \caption{
    Normalised cumulative distribution of the virial parameter with which stars have formed at $1~\Gyr$. Each panel compares the effect of different bulge scale radii at a constant bulge mass, with the columns showing simulations with bulges containing 30, 60 and 90~per~cent of the initial stellar mass, from left to right. Each panel includes a dotted line showing the bulgeless simulation, a black horizontal line indicating the median of the distribution, and a black vertical line indicating virial equilibrium, i.e.\ $\avir=1$.} 
    \label{fig:avir_cumulative_vSFE_stars}
\end{figure*}

The gas that formed stars (bottom row of Figure~\ref{fig:avirradprof_vSFE}) has a lower median $\avir$ than all gas eligible for star formation, as well as a clearer segregation between the different bulges within the inner $2{-}4~\kpc$. The bulgeless and weakest bulge galaxies (noB, B\_M30\_R3 and B\_M30\_R2) form stars with virial parameters in range $\avir=1{-}5$, as expected from star formation theory and in accordance with recent observations \citep[e.g.][]{Sun2018,Schruba2019}. The more bulge dominated galaxies exhibit larger median virial parameters, especially in the inner $1{-}2~\kpc$, reaching a maximum of $\avir=50$ for run B\_M90\_R1. Although this might seem disconcertingly high, such high virial parameters have been observed for some clouds in the CMZ of the Milky Way \citep[e.g.][]{Kruijssen2014b,Kauffmann2017b}. In addition, Figure~\ref{fig:avir_cumulative_vSFE_stars} reveals that the vast majority of stars forms with much smaller virial parameters.
Because $\avir$ is constituted by multiple physical quantities (cloud mass, radius, velocity dispersion) that are all affected by star formation and feedback over time, the median virial parameter exhibits a larger time variation between different snapshots than the quantities it is based on. This results in larger scatter than seen for the velocity dispersions in Figure~\ref{fig:veldispprof_vSFE}.

For the most compact bulges, there is a pronounced increase of the time variability of the virial parameter of gas eligible for star formation at the very centre of the galaxy, with large downward excursions. Over the course of the simulation, gas flows towards the centre of the galaxy. Due to the amount of turbulence induced by the compact bulges, star formation is strongly suppressed and gas continues to build up. Eventually, the gas becomes so dense that it is no longer gravitationally fully resolved, i.e.\ overdensities identified by our model are smaller than the gravitiational softening length, despite consisting of tens of gas cells. Despite its high velocity dispersions, the virial parameter then decreases again, resulting in star formation. As discussed in more detail in Section~\ref{ss:numcavs}, this downward spike of the virial parameter is at least partially a numerical artefact stemming from our resolution, which motivates the exclusion of the central $300~\pc$ from the star formation related analysis (see Section~\ref{ss:vcSFE}).

We show the cumulative distribution of virial parameters for all gas cells that formed stars in Figure~\ref{fig:avir_cumulative_vSFE_stars}. Stars in the bulgeless simulation form from gas with a median virial parameter of $\avir=1.1$. The median virial parameters are higher for the galaxies with bulges, more so for galaxies with more massive and compact bulges, up to a maximum of $\sim 3.5$ for the most bulge-dominated galaxy. This is a reflection of the elevated virial parameters in (the central parts of) these galaxies compared to the bulgeless run, as seen in the bottom row of Figure~\ref{fig:avirradprof_vSFE}. Between $2{-}8$~per~cent of stars are formed with $\avir \geq 10$ across all simulations, and only a handful of stars with $\avir \geq 20$.

\subsection{Turbulent Pressure} \label{ss:vSFE_pturb}
\begin{figure*}
    \centering
    \includegraphics[width=1.\linewidth]{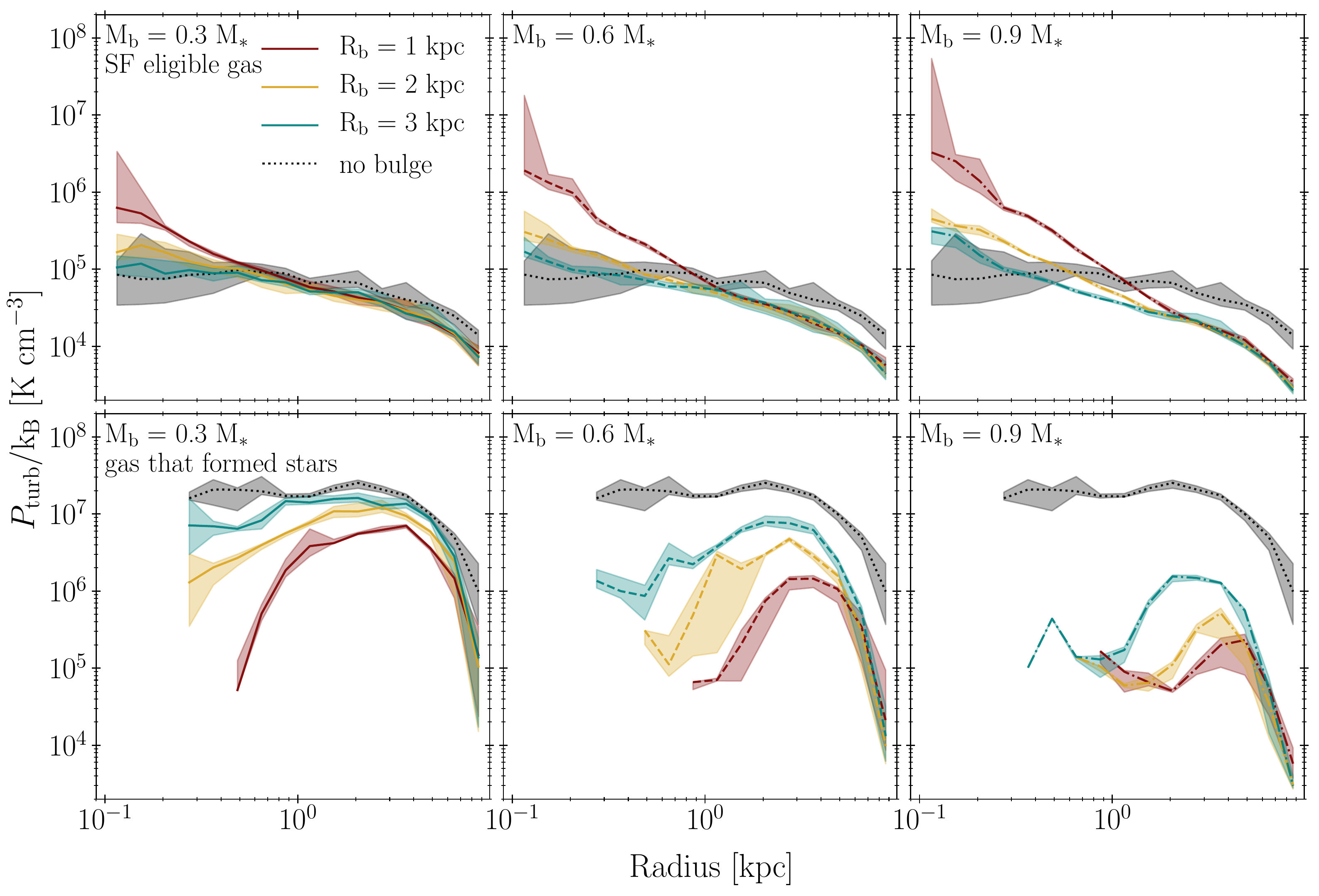}%
    \caption{
    Radial profiles of the turbulent gas pressure for different gravitational potentials. Each panel compares the effect of different bulge scale radii at a constant bulge mass, with the columns showing simulations with bulges containing 30, 60 and 90~per~cent of the initial stellar mass, from left to right. As before, the top row shows the result for gas that satisfies the density and temperature thresholds for star formation, whereas the bottom row only includes gas from which stars formed. The central $300~\pc$ are excluded from the analysis of the bottom panels (see Section~\ref{ss:vcSFE}). Each panel includes a dotted line showing the bulgeless simulation.
    } 
    \label{fig:pturbradprof_vSFE}
\end{figure*}

Figure~\ref{fig:pturbradprof_vSFE} shows the radial profiles of the turbulent gas pressure for different gravitational potentials. The clear differences between the different simulations suggests that turbulent pressure is the most unambiguous tracer of the underlying gravitational potential. The bulgeless galaxy has a roughly constant median $\Pt$ out to $\sim5~\kpc$, before it starts declining, as both the gas density and the velocity dispersion decrease. All of the galaxies with bulges have steeper profiles. They have higher turbulent pressures in their centres, with a faster decline towards the outskirts. Even for the $0.3\Mstar$ bulges, where the central values fall within the uncertainty implied by the time variation of the bulgeless galaxy, we find a distinctly different shape of the pressure profile compared to the bulgeless model. The increase of the median $\Pt$ within the inner kpc is achieved at lower bulge densities than for $\sigma$ and $\avir$, with the B\_M30\_R2 already showing a statistically significant central enhancement of $\Pt$. This results from the fact that increased gas densities and velocity dispersions both contribute to boosting the pressure, such that $\Pt$ is the cleanest tracer of the effect of the bulge on the ISM properties.

Broadly speaking, the trends of $\Pt$ with $\Rb$ and $\Mb$ are similar to the behaviour of $\avir$ and $\sigma$ with radius. The most compact galaxies have median central pressures that are 1-2 orders of magnitude larger than that of the bulgeless galaxy.
This links back to the star formation activity within these galaxies, which we will quantify further in Section~\ref{ss:vSFE}. A dearth of star formation in the central region allows the high density gas to dominate, such that nearly all gas is eligible for star formation. In combination with the higher velocity dispersions, this leads to a consistently high $\Pt$.

As for the velocity dispersion, the strong increase in turbulent pressure towards the centre of the galaxy traces the gravitational potential generated by the bulge. The corresponding increase of the turbulent pressure for larger bulge mass fractions and smaller scale radii is an observationally testable prediction of our simulations.

Contrary to what we find for the radial profiles of the velocity dispersion and virial parameter, the turbulent gas pressure profiles of the gas that formed stars do not closely mirror those of the star formation eligible gas. The shape of the bulgeless galaxy's profile is similar to that of the eligible gas, but median pressures are two orders of magnitude higher. For galaxies with bulges, the trend seen in the top panels is inverted: the more bulge-dominated a galaxy, the lower the median $\Pt$ of the gas that formed stars. This happens because the bulge suppresses the formation of density contrasts in the ISM (see Section~\ref{vSFE_gasSDdistbns}). While the bulgeless galaxy achieves considerable fragmentation and thus major density (and therefore pressure) contrast between the diffuse ISM and its condensations, the suppression of fragmentation by the bulges in the other simulations means that the pressure contrasts are smaller.

\subsection{ISM morphology} \label{vSFE_gasSDdistbns}
\begin{figure*}
    \centering
    \includegraphics[width=1.\linewidth]{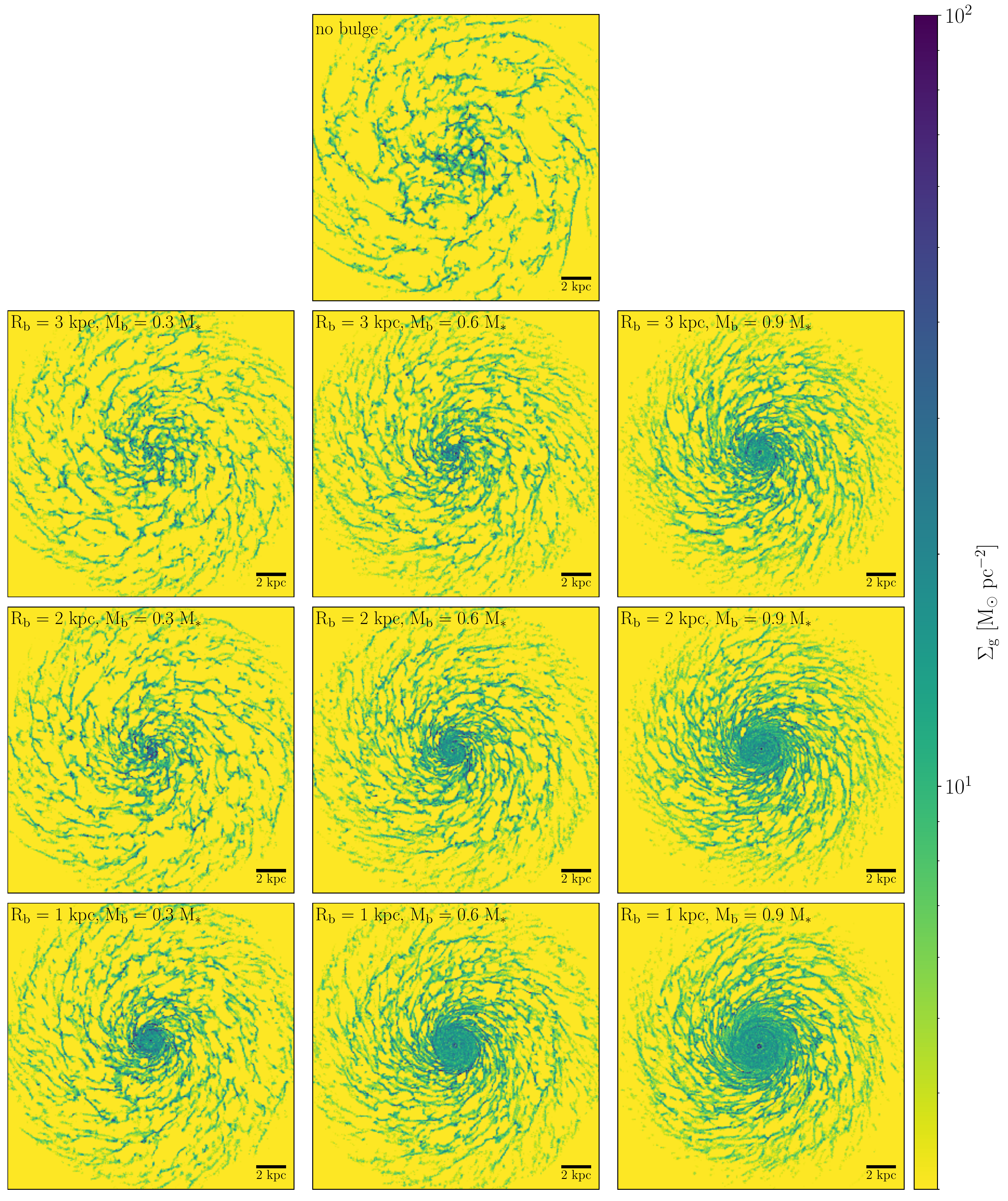}%
    \caption{
    Surface density projection of the gas discs evolved with a dynamics-dependent star formation model. The top panel shows the bulgeless simulation. In the other panels, the bulge mass fraction increases towards the right and the bulge scale radius increases upwards, as indicated by the annotations, such that the bulge density increases towards the bottom right. The maps are shown at 600~Myr after the start of the each simulation and measure $20~\kpc$ on a side. The suppressed SFR in bulge-dominated galaxies enables the build-up of a quiescent, undisturbed, and therefore smooth nuclear gas disc. The spatial extent of this disc increases with the density of the bulge. Only galaxies with a negligible bulge component show signs of substructure driven by gravitational instability and stellar feedback.
    }
    \label{fig:vSFE_sdproj}
\end{figure*}

Lastly, we focus on how the gravitational potential affects the spatial distribution and structure of the ISM within the galaxies. Figure~\ref{fig:vSFE_sdproj} shows the projected gas surface density maps of all galaxies evolved with a dynamics-dependent SFE at $600\Myr$ after the start of the simulation. Each galaxy starts out with the same, smooth exponential gas disc. Therefore any difference in disc structure is due to the combined effects of star formation, stellar feedback, and the dynamical evolution in the underlying gravitational potential, which differs between the simulations depending on the bulge properties.

The outskirts of all galaxies look similar, with dense gas arrayed along tightly wound spiral arms. However, the centres clearly differ between the ten simulations shown here. In the most massive and compact bulges, the gas settles into a smooth, dense disc. The spatial extent of these discs range from being several hundred pc in radius for the intermediate bulges (B\_M30\_R1, B\_M60\_R2, and B\_M90\_R3) to extending a size of $2~\kpc$ for the compact bulge simulation (B\_M90\_R1). By contrast, the bulgeless galaxy and the simulation with the weakest bulge (B\_M30\_R3) host a central gas reservoir with considerable substructure, even though they also contain a large amount of dense gas in their central region. These central regions do not resemble discs, but fragment into star-forming units and are subsequently disturbed further by stellar feedback.

The gas surface density maps of Figure~\ref{fig:vSFE_sdproj} show a stellar bulge stabilises the ISM of its host galaxy. Thanks to the increased shear velocities, the gas velocity dispersions are elevated, which together prevents the gravitational instability of the gas reservoir and suppresses star formation, in the process preventing the disruption of the ISM by stellar feedback. This is a powerful illustration of the interplay between galactic dynamics, ISM structure, and star formation, which predicts clear, monotonic trends with the morphology of the host galaxy.

\subsection{Star Formation} \label{ss:vSFE}
\begin{figure*}
    \centering
    \includegraphics[width=1.\linewidth]{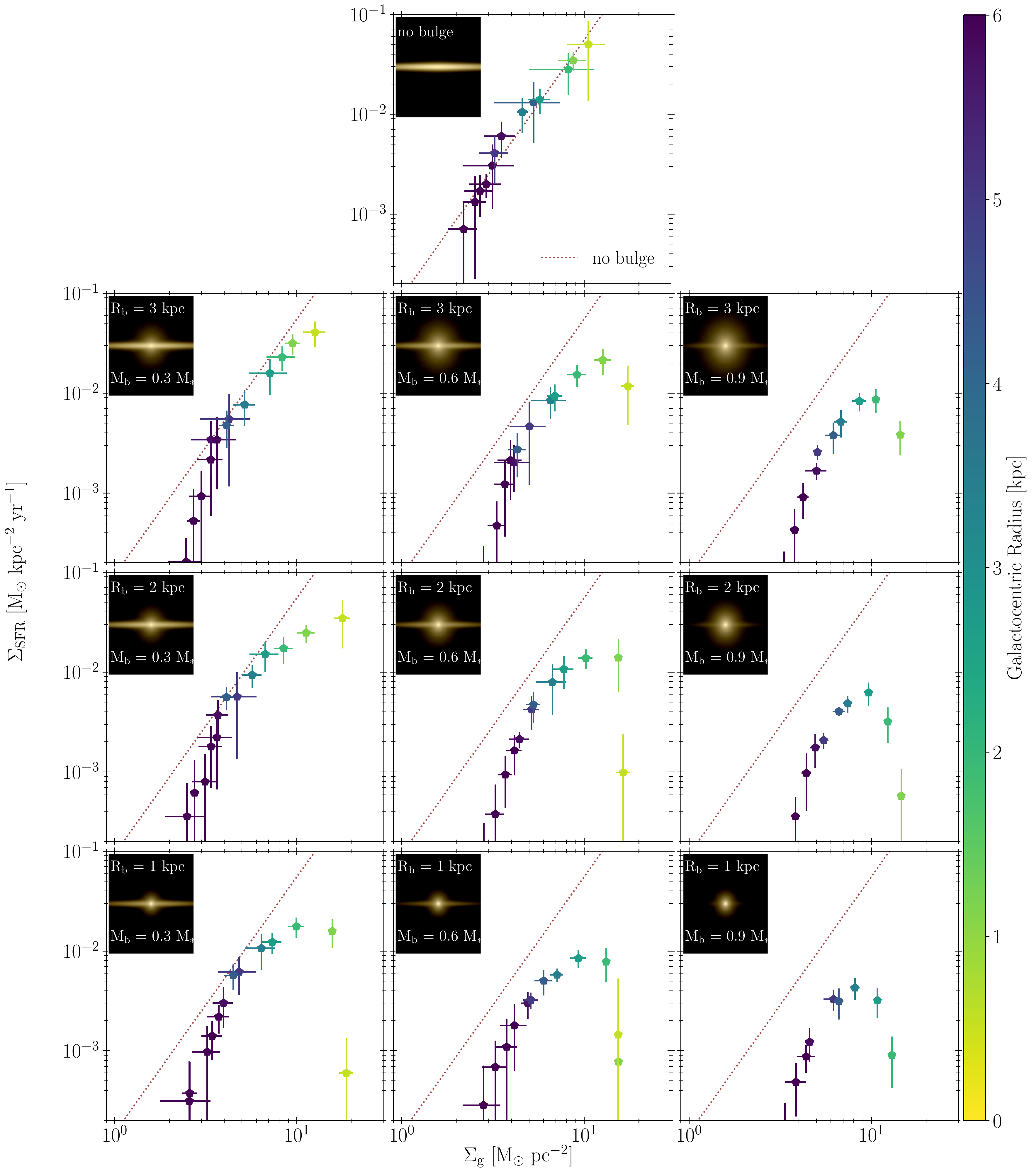}%
    \caption{
    SFR surface density as a function of gas surface density for all galaxies evolved using a dynamics-dependent $\eff$. The colour coding indicates the galactocentric radius. For reference, the top-left corner of each panel shows a fake colour image of the stellar content of each galaxy. The dotted red line indicates the best-fitting power law to the star formation relation of the bulgeless galaxy, and is included to guide the eye and highlight how the star formation relation changes due to differences in the gravitational potential. The SFRs are calculated from stars that formed within the past 10~$\Myr$ of any snapshot. Each point represents a time average of snapshots separated by $100~\Myr$; the snapshot-to-snapshot variation is shown by the error bars. The central $300~\pc$ are excluded from the analysis (see Section~\ref{ss:vcSFE}). This figure shows that the presence of a spheroid suppresses the SFR towards small galactocentric radii in all simulations, and that this suppression intensifies for higher-density bulges, which are located towards the bottom right.
    }
    \label{fig:KS_vSFE}
\end{figure*}

We conclude our parameter survey of the impact of the gravitational potential on the ISM and star formation by focusing on the SFR of the simulated galaxies. Based on the results presented in Sections~\ref{ss:vSFE_veldisp}-\ref{ss:vSFE_pturb}, with the gas velocity dispersion, virial parameter, and turbulent pressure all increasing towards the galactic centres, and most strongly so for the most massive and compact bulges, we expect the SFR in these galaxies to follow suit. Figure~\ref{fig:KS_vSFE} shows the star formation relation, between the SFR surface density and the gas surface density, for all galaxies in the sample. The data points and colours indicate different galactocentric radii, showing how the star formation relation depends on radius. As elsewhere in this section, we only consider the simulations run with a dynamics-dependent star formation model. As before, we exclude the central $300~\pc$ from our analysis, to ensure that our conclusions are not influenced by unresolved star formation at the centre. This is discussed further in Section~\ref{ss:numcavs}.

The star formation relation follows a simple, monotonic, power law form for the bulgeless galaxy (see the red dotted lines in Figure~\ref{fig:KS_vSFE}). However, this changes as the bulge mass fraction and compactness increases, particularly towards the centres of the galaxies. Out of all galaxies, only the weakest bulge simulation (B\_M30\_R3) is marginally consistent with the bulgeless galaxy to within the time variability between different snapshots, as encapsulated by the error bars. All other galaxies exhibit a pronounced flattening or turnover of the star formation relation towards the galactic centre. The degree of the flattening and the overall suppression of the SFR within the galaxy increase for more massive and compact bulges. For the galaxy with the most compact bulge (B\_M90\_R1), this yields a total, galaxy-wide SFR that is a factor of $\sim5$ lower than that of the bulgeless galaxy.

\begin{figure*}
    \centering
    \includegraphics[width=.47\linewidth]{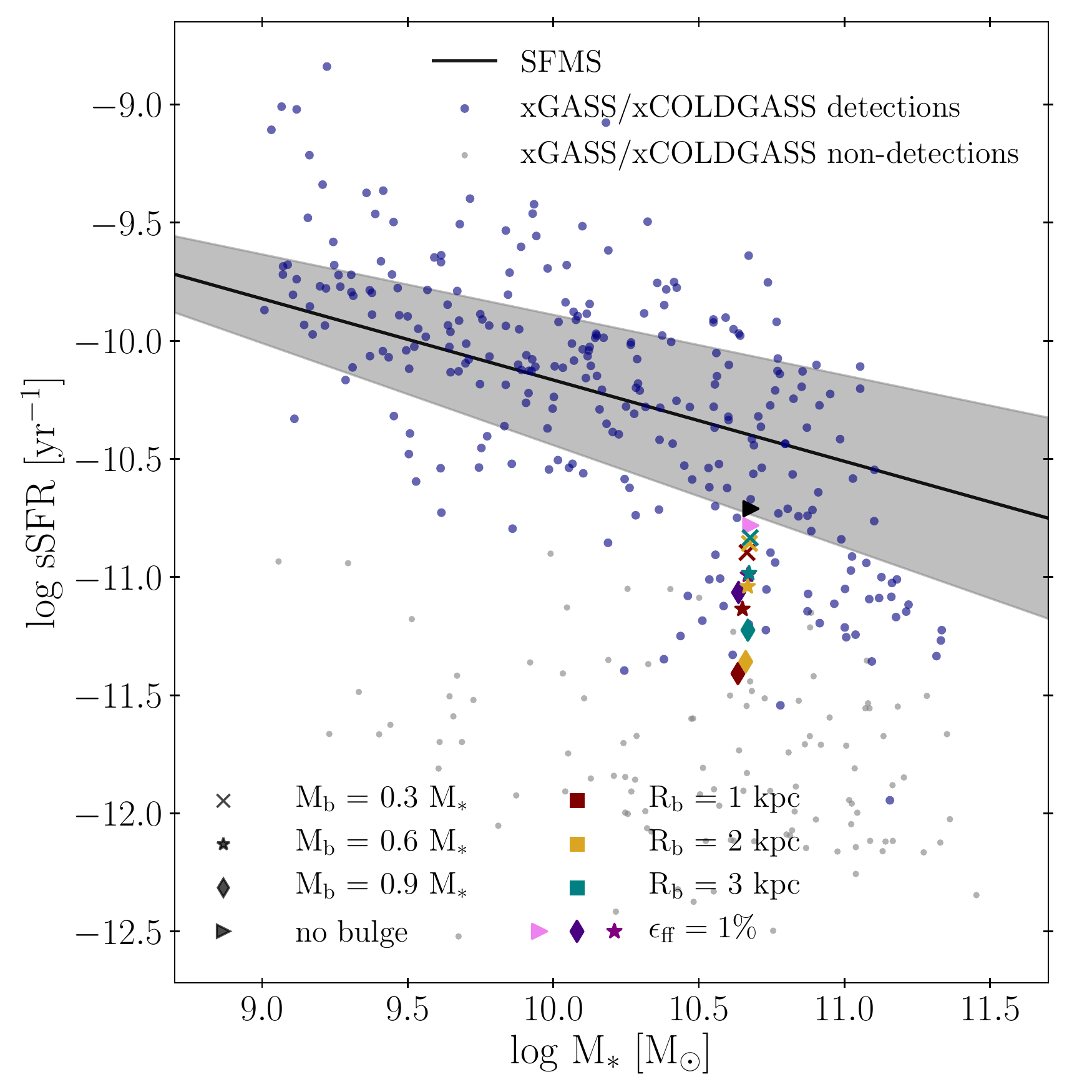}%
    \includegraphics[width=.47\linewidth]{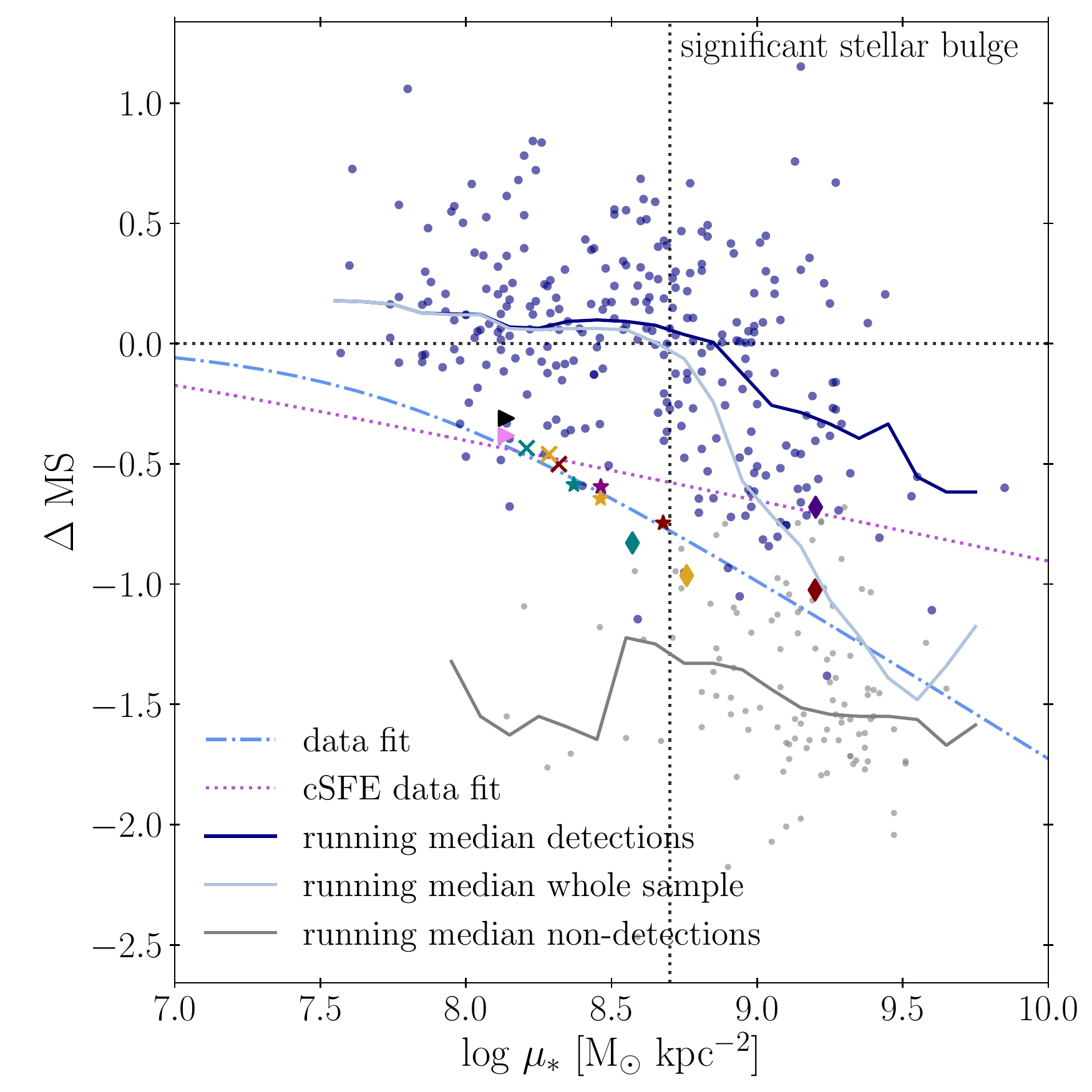}%
    \caption{Simulation data overplotted on the data from the xCOLDGASS \protect\citep{Saintonge2017} and the xGASS \protect\citep{Catinella2018} surveys. The left panel shows the sSFR--stellar mass plane and the right panel shows the offset from the main sequence as a function of stellar surface density ($\mus$). 
    The definition of the star formation main sequence is taken from \protect\citet{Catinella2018}. 
    All simulated galaxies with dynamics-dependent star formation exhibit a strong trend: the global SFR falls further below the main sequence, the more bulge dominated the galaxy. This trend is shown by the dot-dashed blue line in the right panel; solid lines show the running median of the detected (navy), non-detected (grey) and entire sample (steel blue) of xGASS/xCOLDGASS galaxies. Constant efficiency galaxies follow a much shallower trend with $\mus$, as indicated by the dotted purple line. As for the Schmidt-Kennicutt diagrams the central $300~\pc$ of the simulated galaxies are excluded from the analysis.}
    \label{fig:SFMS_and_deltaMS}
\end{figure*}

All galaxies exhibit a clear central suppression of the SFR, by up to two orders of magnitude (which is similar to the suppression of the SFR observed in the CMZ of the Milky Way, \citealt{Longmore2013}). However, the uncertainty introduced by the time variability of the SFR also seems to increase as the suppression becomes more pronounced, towards more dominant bulges. This is a statistical effect. We use radial bins with a constant radial thickness, implying that the area over which the SFR is integrated for each bin increases towards the outskirts of the galaxy, driving down the uncertainties. However, outside radii of $\sim3$~kpc, the scatter increases again. This results from the outward decrease of the gas surface density and the SFR, implying fewer star formation events at large radii and increased stochasticity. The time variability is the smallest at radii of a few kpc.

In conclusion, Figure~\ref{fig:KS_vSFE} shows that the gravitational potential generated by a stellar spheroid strongly influences the star formation relation of galaxies. The galaxies with the most massive and compact bulges are affected the most, showing a flattening or turnover of the star formation relation, as well as the lowest total SFR with suppressions by up to a factor of $\sim5$.

\section{Discussion} \label{s:Dis}

\subsection{Comparison with observations} \label{ss:obscomp}
The discussion of Sections~\ref{vSFE_gasSDdistbns} and~\ref{ss:vSFE} shows that the presence of a bulge leads to decreased ISM fragmentation and suppressed SFRs towards galactic centres. Here, we place these results in the context of the observed galaxy population.

\subsubsection{Main sequence of star-forming galaxies}
It was shown in Section~\ref{ss:vSFE} that the integrated SFRs of the simulated galaxies can be suppressed by up to a factor of $\sim5$ due to the dynamical effects induced by a stellar spheroid. In Figure~\ref{fig:SFMS_and_deltaMS}, we place the implications of this result in the context of the galaxy population. We show the global sSFR as a function of stellar mass (spanning the `main sequence of star-forming galaxies', e.g.\ \citealt{Noeske2007,Peng2010}), as well as the offset of the main sequence as a function of stellar surface density ($\mus$), both for our suite of simulations and for the galaxy population observed as part of the xCOLDGASS \citep{Saintonge2017} and xGASS \citep{Catinella2018} surveys. Before proceeding, we note that all of our galaxies have the same initial $\Mstar$ and low gas fraction of 5~per~cent. Being isolated simulations, they do not include gas accretion or any other influences of the large-scale cosmic environment and are chosen to resemble $z=0$ galaxies.

The left-hand panel of Figure~\ref{fig:SFMS_and_deltaMS} shows that the global sSFR of our sample ranges from the star formation main sequence (for the bulgeless galaxy), to the lowest sSFR at which gas is detected within the surveyed galaxies, approximately 1~dex below the main sequence, where star formation in galaxies has been `quenched'. The simulated galaxies thus span a range consistent with the observed, vertical scatter below the star formation main sequence. This means that, even without including any feedback from active galactic nuclei, our simulations bridge the transition region (the `green valley') towards the quenched galaxy population (the `red cloud'), purely relying on the interplay between galactic morphology, (hydro)dynamics, the properties of the ISM, and star formation.

It is reasonable to expect that the offset from the main sequence depends on the density of the stellar spheroid or bulge. The right-hand panel of Figure~\ref{fig:SFMS_and_deltaMS} shows the offset from the main sequence as a function of the stellar mass surface density ($\mus$), together with the observed galaxies from xCOLDGASS and xGASS. Our simulations with a dynamics-dependent star formation model follow a clear trend of increased quenching of star formation towards increasing $\mus$. This highlights that the stellar surface density (i.e.\ a combination of the bulge mass and scale radius) is the physical quantity driving the suppression of star formation. Compared to the running medians of detections and non-detections and the full sample, the fit to our sample does not quantitatively match any of the individual trends and also exhibits considerably less scatter than the observed galaxies. This is not necessarily surprising. Firstly, we did not fine-tune our simulations to match the xGASS and xCOLDGASS galaxies. Secondly, we only vary the bulge properties in this experiment, but it is likely that other quantities (e.g.\ gas fraction) affect the offset from the main sequence. We will explore this further in a follow-up study (Gensior et al.\ in prep.).

Most importantly, our simulations reproduce the overall trend that star formation is suppressed or quenched as the stellar surface density of the central bulge or spheroid increases. By contrast, the simulations with a constant SFE show a very different trend. Their offset from the main sequence is considerably smaller, with a much shallower dependence on $\mus$. The total dynamic range in sSFR is a factor of $\sim2$, much smaller than for the dynamics-dependent SFE. Finally, Figure~\ref{fig:KS_vSFE} illustrates that the bulge-dominated galaxies in our sample also qualitatively reproduce the trends observed by \cite{MendezAbreu2019}. The more bulge-dominated they are, the more significant the drop of SFR towards the centre. 

\subsubsection{ISM properties}
A clear prediction of our models is that the velocity dispersion, virial parameter, and ISM pressure increase towards galactic centres. For velocity dispersions and ISM pressures, this is a well-established observational result \citep[e.g.][]{Kruijssen2013,Leroy2016,Sun2018,Sun2020,Schruba2019}. Indeed, the velocity dispersions spanned by our simulations are consistent with the range found in the above studies. The turbulent pressures are also consistent with observations of star-forming galaxies in the local Universe \citep[e.g.][]{Faesi2018}. There is tentative observational evidence that the gas virial parameter also increases towards galactic centres, especially in the presence of strong dynamical features \citep{Sun2018}. The radial median values of the virial parameter found in our simulations are 10--100, which might seem higher than expected, but falls within the range that \cite{Sun2018} find for M33 and M31.

A number of early-type galaxies are observed to host molecular gas in an almost featureless, smooth disc of molecular gas around the galactic centre \citep[e.g.][]{North2019}. Similar, smooth gas discs form naturally in our simulations with a dynamics-dependent SFE, with a spatial extent that increases with the stellar (surface) density of the bulge or spheroid. These smooth gas discs do not form when a constant SFE is used, in which case considerable substructure remains. This shows that the discovery of smooth gas discs in galaxy spheroids supports the idea that the SFE depends on the dynamical state of the ISM. At first sight, the gas in our galaxies extends further than the inner $1-2\kpc$ (see e.g.\ Figure~\ref{fig:vSFE_sdproj}), as found in the observations of \citet{North2019}. However, we remind the reader that we model all gas within the galaxies, without applying any density threshold or carrying out synthetic observations. Given the gas surface density profiles of our simulations (see Figure~\ref{fig:vcSFE_sd_radprof}), it is likely that the molecular gas traced by CO drops off beyond radii of $\sim2~\kpc$.

In summary, the ISM properties of our simulations are qualitatively consistent with those found in observations, and provide an interesting physical perspective on the existence of quiescent, smooth gas reservoirs in early-type galaxies. We expect that quantitative comparisons between these simulations and the observations can help shed more light on the detailed interplay between dynamics, ISM structure, and star formation.

\subsection{Comparison with other simulations} \label{ss:numcomp}
In Section~\ref{s:intro}, we summarised a variety of simulations in the literature focusing on the impact of galaxy morphology on star formation. To date, the main missing step in the `morphological quenching' framework had been how galactic morphology translates into a cloud-scale suppression of star formation. In our simulations, we capture this process by resolving the cold ISM and including a dynamics-dependent sub-grid model for star formation, in which star formation is gradually suppressed towards high virial parameters. We show that the inclusion of such a model unlocks a variety of interesting predictions that may explain the observed correlations between galactic morphology, star formation, and quenching. Our suite of simulations systematically surveys the parameter space spanned by the properties of the stellar spheroids. However, our results must be regarded as a single, homogeneous set of experiments, using a single numerical method and basing itself on a number of necessary numerical choices. Here we discuss our results in the context of other numerical studies.

In a recent study, \cite{Su2019} find that morphological quenching is not capable of shutting down star formation completely, even if the SFR is lower in the presence of a bulge. As Figures~\ref{fig:KS_vSFE} and~\ref{fig:SFMS_and_deltaMS} show, we find a similar trend for the galaxies in our suite, in which the SFR of the most bulge-dominated galaxy (B\_M90\_R1) is the lowest. However, the difference in SFR between our bulgeless and most bulge-dominated galaxy is $\sim3$ times larger than found by \citet{Su2019}. However, there are a number of differences in the numerical setup and the initial condition that could be responsible for this offset. \citet{Su2019} use the star formation model from \textsc{fire2} \citep{Hopkins2018b}, in which star formation can only proceed in self-gravitating regions, with a constant $\eff=1$, whereas our simulations use a gradual suppression of the SFR as a smooth function of the virial parameter. Furthermore, the isolated galaxy simulations of \citet{Su2019} include a halo of hot gas, which by cooling feeds the supply of star-forming gas, boosting gas fraction and star formation. This likely contributes to the larger SFRs relative to the ones in our galaxy sample. 

Our simulations with a constant SFE are similar to the spiral and elliptical galaxy simulations with a gas fraction of 4.5~per~cent from \citet{Martig2013}. We find a smaller difference in global SFR between noB\_cSFE and B\_M60\_R2\_cSFE/B\_M90\_R1\_cSFE compared to their simulations, and a slightly larger normalisation than their star formation relation. However, the agreement is satisfactory overall. Our simulations show that the dependence of the sub-grid star formation model on galactic dynamics is required to reproduce a more pronounced suppression of star formation and change the star formation relation. Our findings are also in good agreement with the early-type galaxy simulation of \cite{Kretschmer2019}, which shows that a bulge-dominated spheroid can suppress star formation in the centre of a galaxy. Their study also further supports the argument that a more physically-motivated, dynamics-dependent star formation model is necessary to accurately model star formation in such systems.

Even if they do not use it to investigate morphological quenching, \cite{Semenov2016,Semenov2017,Semenov2018} make use of the \citet{Padoan2012} parametrisation of the virial parameter to calculate the SFE in their simulations. Apart from using a different method for calculating $\avir$, their sub-grid star formation model is the most similar in the literature compared to the prescription we are using. Despite the fact that their isolated galaxy contains a bulge, \cite{Semenov2016} do not see any central suppression of star formation. However, their simulation differs in two important ways from the galaxies in our suite. Firstly, their bulge component only holds $\sim 9$ per cent of the initial stellar mass of the galaxy confined to the centre, thus being most similar to our bulgeless galaxy, which does not significantly deviate from the observed star formation relation \citep{Kennicutt2012}. Secondly, the gas content of their simulation is a factor of 4 larger, at 20~per~cent of the initial stellar mass, compared to the 5~per~cent that we use to mimic gas-poor, early-type galaxies.

Previous work has suggested that there may exist a critical gas fraction below which `morphological quenching' (or, in the context of our results, rather the `dynamical suppression of star formation') may proceed. \cite{Martig2013} perform simulations of idealised galaxy models and find a considerably stronger suppression of star formation at a gas fraction of 1.3~per~cent than at 4.5~per~cent. In the context of our work, we remind the reader that \cite{Martig2013} use a constant star formation efficiency, which implies that the dynamical suppression of star formation is less pronounced than when using a dynamics-dependent model. As a result, this suppression may not be able to manifest itself at gas fractions as high as in our simulations. None the less, the important insight is that the suppression weakens towards larger gas fractions. The high-redshift ($z>5.7$) simulations of highly gas-rich ($\sim50$~per~cent) galaxies by \citet{Trebitsch2017} further highlight that gas fraction is an important quantity to consider, because bursty star formation continues to proceed at a roughly constant time-averaged rate in their most massive halo, even after the build-up of a small central bulge. Similarly, \cite{Kretschmer2019} only see a suppression of star formation in their cosmological zoom-in simulation of an early-type galaxy after a drop in gas fraction following a merger-induced starburst, once again indicating that gas fraction plays a key role in the suppression of star formation by a dominant bulge component. Exploring how the gas fraction of our galaxies affects their SFRs is beyond the scope of this paper, but we are preparing a follow-up study in which this is investigated (Gensior et al.\ in prep.).

\subsection{Numerical caveats} \label{ss:numcavs}
\subsubsection{Gravitational softening and the implications for our sub-grid model}
Current numerical simulations of galaxies use particles with masses much larger than those of the individual stars that they are constituted by. To prevent spurious heating from close encounters by these massive particles, gravitational forces must be softened. While adaptive gravitational softening minimises the error in the force calculation \citep{Price2007}, one still has to choose a (minimum) length scale below which the gravitational force is softened. 

Our sub-grid star formation model is aimed at self-consistently identifying an overdense region around a gas cell, for which the gas density and velocity dispersion can then be calculated. Since the gravitational softening length is larger than the cell radius by definition, it is conceivable that in a very dense, clumpy medium the overdensity identified is smaller or comparable in radius to the softening length. We have quantified this directly in our simulations and find that this occurs for $\leq 1$~per~cent of gas cells in our simulation at any given timestep, where the exact fraction depends on the prior amount of star formation and feedback experienced by the gas. The virial parameters corresponding to these regions are all $\avir < 2$, with the majority having $\avir < 1$.

There are two possible ways in which these edge cases can be handled. Firstly, one can define a hard threshold for $\ltw$, requiring it to be at least equal to the softening length of the cell. Secondly, one can simply accept that some overdensities fall within the gravitationally softened volume. The former option goes against the principle of self-consistently determining an overdensity based on which all other quantities are calculated. Choosing the latter option implies that a disproportionally large fraction of stars will form from overdensities in gravitationally softened regions, as $\eff \propto \exp(-\avir^{0.5})$. 

The purpose of gravitational softening is to prevent numerical noise from artificial hard scatterings. Hence, it can be argued that gas that becomes dense enough to fall within the smallest softening considered should, in the context of this simulation, count as self-gravitating and form stars. This is the standard assumption made when using simple sub-grid star formation models with a constant $\eff$. In that case, a fraction of gas above the star formation threshold (usually some factor less than the maximum gravitationally resolved density) is converted to stars every timestep. However, the main advantage of our model is that the SFE is now determined by gas dynamics, rather than just converting the most gravitationally unresolved gas into stars as models with a constant SFE do. This enables star formation in self-gravitating clumps that might be less dense (i.e.\ that would otherwise have a much lower probability of star formation) and, most importantly, prevents star formation in dense gas that is highly unbound due to external forces acting on the ISM. Both of these features of the dynamical SFE model cannot be achieved by a model with a constant SFE. In the end, only a small subset of gas cells in our model are gravitationally unresolved. We therefore accept this as a minor drawback -- treating this subset as self-gravitating and (potentially) star-forming clouds is equivalent to how they would be treated in a constant SFE model. In light of the fact that we use a more physically-motivated sub-grid star formation model, and to conserve the full self-consistency of the calculation, we choose to not enforce a hard limit for the tree-walk length. 

The above problem is intricately linked to our choice of excluding the central 300~pc from our SFR analysis. Since we do not include any form of feedback other than type~II SNe, gas is able to accumulate at the centres of galaxies, especially in those with high central stellar surface densities. There, the gas is turbulent enough to suppress star formation, resulting in increasing gas densities in the centre, eventually allowing some individual clumps to become smaller than the gravitational softening length. Given the we do not resolve the dynamical mechanisms relevant near galactic centres, which require sub-pc resolution \citep[e.g.][]{Kruijssen2019b}, and also omit feedback from active galactic nuclei, which may also affect the SFE of the most central gas reservoir, we choose to omit the central regions within a 300~pc radius when analysing the SFR. With this radial cut, the majority of stars from gas cells with $\ltw$ smaller than the gravitational softening length are excluded during the analysis. 

\subsubsection{A note on numerical chaos} \label{sss:chaos}
As recently demonstrated by \cite{Keller2019} and \cite{Genel2019}, numerical simulations of galaxy formation and evolution are sensitive to (chaotic) stochasticity. As a means of assessing whether differences between models reflect the underlying differences in physical models or initial conditions, \cite{Keller2019} advise to run a large number of identical simulations with different random seeds. While it is not feasible to simulate every galaxy in our suite a large number of times, all quantities analysed in this paper are represented as averages over time, with an indication of the variance given by error bars. Because we model idealised galaxies, these temporal variation is expected to be analogous to the stochasticity investigated by \cite{Keller2019} and \cite{Genel2019}. Using the temporal variation to quantify the uncertainties on the quantities considered thus ensures that our conclusions remain robust and appropriately reflect how the results depend on the underlying physics.

To further test the effect of numerical effects, we simulated a subset of the galaxies in our suite with different random number seeds (which influence the stochasticity of star formation and thus the subsequent evolution of the galaxy), finding that any differences caused by these numerical effects are smaller than the variation quantities exhibit over time. 

\section{Summary and conclusions} \label{s:SC}

In this work, we present hydrodynamical simulations of a suite of isolated galaxies with gas-fractions appropriate for $z=0$ galaxies, to systematically explore whether stellar bulges or spheroids can dynamically suppress star formation. We systematically vary the bulge mass fraction and scale radius to cover a total of ten different galactic morphologies. We also introduce a new, physically-motivated sub-grid model for star formation that includes a dependence of the SFE per free-fall time ($\eff$) on galactic dynamics via the virial parameter, motivated by the simulations of turbulent fragmentation by \citet{Padoan2012}. Contrary to many previous sub-grid star formation models of this nature, we do not use a sub-grid turbulence model. Instead, the physical quantities entering the virial parameter are calculated by iterating over neighbouring gas cells until an overdensity is identified self-consistently. This approach is analogous to on-the-fly cloud identification and thus establishes the virial parameter for spatially-resolved, physically meaningful units. To enable a comparison to traditional sub-grid star formation models, we also consider simulations with a constant $\eff=1$~per~cent for a subset of the galaxies. These include a bulgeless galaxy and two models with a bulge.

Concerning the two different sub-grid star formation models, we draw the following conclusions:
\begin{enumerate}[leftmargin=0.5cm]
    \item The global SFR exhibits little difference between using a constant or dynamics-dependent SFE, as long as no dominant stellar spheroid is present. However, in the presence of a spheroid, the SFRs of simulations with a dynamics-dependent SFE fall below those with a constant SFE. The more bulge-dominated the galaxy, the larger the difference in SFR.
    
    \item The observed suppression of star formation at the centres of spheroid-dominated galaxies can only be reproduced with the dynamics-dependent sub-grid model. 
    
    \item The gas reservoirs hosted in the centres of galaxies modelled with a dynamics-dependent $\eff$ self-consistently settle into smooth disc of molecular gas. The central ISM of galaxies modelled with a constant SFE remains much substructured and porous. This means that the observed existence of smooth gas discs in early-type galaxies can only be understood if the SFE depends on the dynamical state of the gas.
    
    \item When no bulge is present, there are small differences in gas distribution and ISM kinematics between the sub-grid star formation models, but the galaxies evolve very similarly. 
    
    \item The gas velocity dispersion, virial parameter, and turbulent pressure differ between both sub-grid star formation models, but mostly within the bulge-dominated region. They show no major differences at larger radii.
\end{enumerate}

We proceed to investigate the impact of a stellar spheroid on the SFR and ISM properties of the simulated galaxies. To do so, we adopt the dynamics-dependent star formation model and simulate a suite of ten isolated galaxies that systematically surveys a two-dimensional parameter space of bulge properties. This space spans a factor of three in bulge mass (from 30--90~per~cent of the initial stellar mass in the bulge component) and also in bulge scale radius (from 1--3~kpc). We also include a bulgeless galaxy, so that the final suite ranges from disc-dominated to spheroid-dominated galaxies. All galaxies have a gas fraction of 5~per~cent, appropriate for Milky Way-mass galaxies at $z=0$ \citep[e.g.][]{Saintonge2017,Catinella2018}. Our results can be summarised as follows.
\begin{enumerate}[leftmargin=0.5cm]
    \item Spheroids drive turbulence and increase the gas velocity dispersion, virial parameter, and turbulent pressure towards the galactic centre. The strengths of these effects depends on the bulge mass and radius. More compact (smaller $\Rb$) and more massive (larger $\Mb$) bulges tend to drive more turbulence. The radial range across which the ISM properties change is the largest in galaxies with the most massive bulges.
    
    \item As a result of the above changes to the ISM, a stellar spheroid stabilises the ISM of its host galaxy. Thanks to the increased shear velocities, the gas velocity dispersions are elevated, which together prevents the gravitational instability of the gas reservoir and suppresses fragmention, thereby also preventing the disruption of the ISM by stellar feedback. This leads to the build-up of a smooth disc of molecular gas around the centre of the galaxy. The smoothness and spatial extent of this disc increases for more bulge-dominated galaxies.
    
    \item As a result of the high virial state and the suppression of fragmentation, star formation is dynamically suppressed in the centres of the simulated galaxies hosting a bulge component. This leads to a flattening of the spatially-resolved star formation relations in these systems, because the gas surface density typically increases inwards.
    
    \item In bulge-dominated systems, the dynamical suppression star formation in the centres leads to a measurable galaxy-integrated decrease of the SFR. In our simulations, the factor by which the global SFR is suppressed is up to a factor of $\sim5$. This drives an offset from the main sequence of star forming galaxies. As a result, galaxies with dominant spheroids populate the green valley and reach into the red cloud of quenched galaxies. This offset from the main sequence increases towards larger stellar surface densities of the bulge, clearly highlighting the impact of the gravitational potential.
    
    \item We argue that the `dynamical suppression' of the SFR in galaxy spheroids\footnote{This terminology is preferred over `morphological quenching', which skips the dynamical step that we find to be crucial.} is capable of significantly affecting the global SFR and the subsequent evolution of galaxies, presumably most strongly for those with low ($\la5$~per~cent) gas fractions. This form of `quenching' proceeds without expelling the gas by feedback from massive stars or active galactic nuclei, but by dynamically stabilising the gas reservoir and rendering it quiescent.
    
    \item Following from these results, the SFR of galaxies is not exclusively set by the balance between accretion and feedback, but carries an important dependence on the gravitational potential. This dependence can dominate at the low gas fractions that characterise early-type galaxies and the bulges of late-type galaxies.
\end{enumerate}

We demonstrate that the gravitational potential plays an important role in regulating star formation in spheroid-dominated galaxies, and is capable of dynamically quenching star formation in these systems. Including a dependence on gas dynamics in the sub-grid star formation model is crucial to fully model this effect in simulations of galaxy formation and evolution. However, as the comparison of our simulations to observations shows, even in spheroid-dominated systems there exists considerable scatter around the relation between decreasing SFR with the stellar surface density of the spheroid. This implies that additional physical quantities, such as the gas fraction, are similarly relevant in setting the galactic SFR. This leaves important avenues for extending our analysis and improving the predictive power of these simulations.

\section*{Acknowledgements}
We thank Volker Springel for allowing us access to Arepo. We thank an anonymous referee for a helpful and constructive report, and the members of the MUSTANG group (M{\'e}lanie Chevance, Daniel Haydon, Alexander Hygate, Sarah Jeffreson, Jaeyeon Kim, Maya Petkova, Marta Reina-Campos, Sebastian Trujillo-Gomez, and Jacob Ward) for for helpful discussions. JG would like to thank the participants of the conference The Life and Death of Star-Forming Galaxies, especially Timothy Davis, Eric Emsellem and Richard McDermid, for fruitful discussions. The authors acknowledge support by the High Performance and Cloud Computing Group at the Zentrum f{\"u}r Datenverarbeitung of the University of T{\"u}bingen, the state of Baden-W{\"u}rttemberg through bwHPC and the German Research Foundation (DFG) through grant no INST 37/935-1 FUGG. JG and JMDK gratefully acknowledge funding from the Deutsche Forschungsgemeinschaft (DFG, German Research Foundation) through an Emmy Noether Research Group (grant number KR4801/1-1). JMDK gratefully acknowledges funding from the DFG Sachbeihilfe (grant number KR4801/2-1). JMDK and BWK gratefully acknowledge funding from the European Research Council (ERC) under the European Union's Horizon 2020 research and innovation programme via the ERC Starting Grant MUSTANG (grant agreement number 714907). BWK acknowledges funding in the form of a Postdoctoral Research Fellowship from the Alexander von Humboldt Stiftung.




\bibliographystyle{mnras}
\bibliography{HoD_refs}



\appendix

\section{Resolution test} \label{app:ResTest}
\begin{figure}
    \centering
    \includegraphics[width=0.95\linewidth]{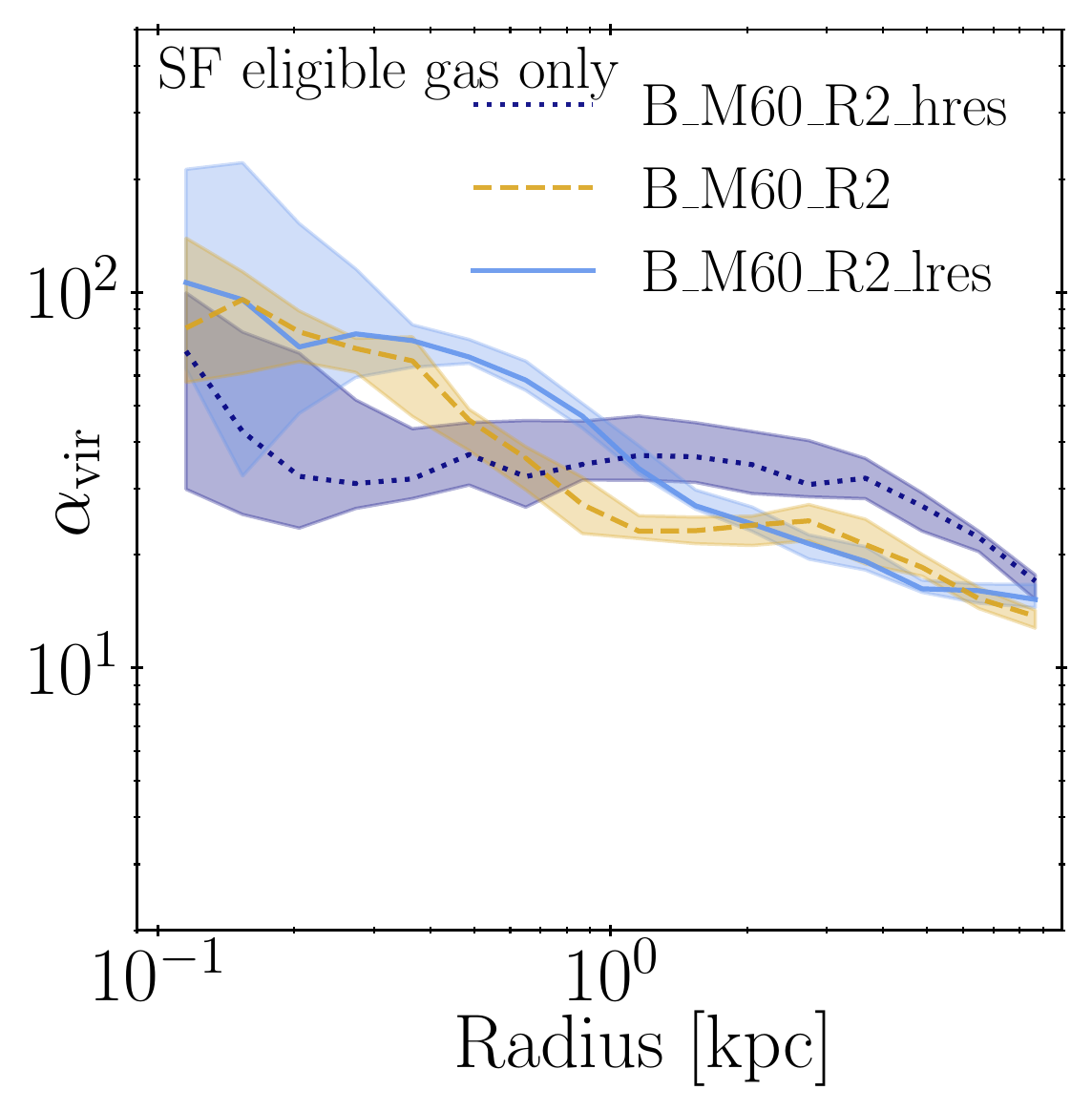}%
    \caption{Time-averaged radial profiles of the virial parameter, shown for three simulations of the fiducial bulge galaxy spanning a decade in numerical resolution. All profiles follow a similar trend, but differences persist. These differences result from stellar feedback having a slightly different effect at different resolutions.}
    \label{fig:radprof_avir_diffres}
\end{figure}

The dynamics-dependent sub-grid star formation model includes a weak dependence on the (number of) neighbouring gas cells during the initial tree-walk. Therefore, we simulate the fiducial initial conditions at both higher and lower resolution (runs B\_M60\_R2\_hres and B\_M60\_R2\_lres respectively) to test the convergence of the star formation model over a decade in resolution. The virial parameter is the crucial quantity for which the effect of resolution has to be determined, because it sets the SFE. In Figure~\ref{fig:radprof_avir_diffres}, we show the time-averaged median virial parameter as a function of the galactocentric radius. It shows that the virial parameter follows the same radial trends despite the different mass resolution.
With feedback also being affected by resolution, some offsets and slight variation in profile shape are expected. We therefore consider the convergence expressed by Figure~\ref{fig:radprof_avir_diffres} to be satisfactory.

\begin{figure}
    \centering
    \includegraphics[width=0.95\linewidth]{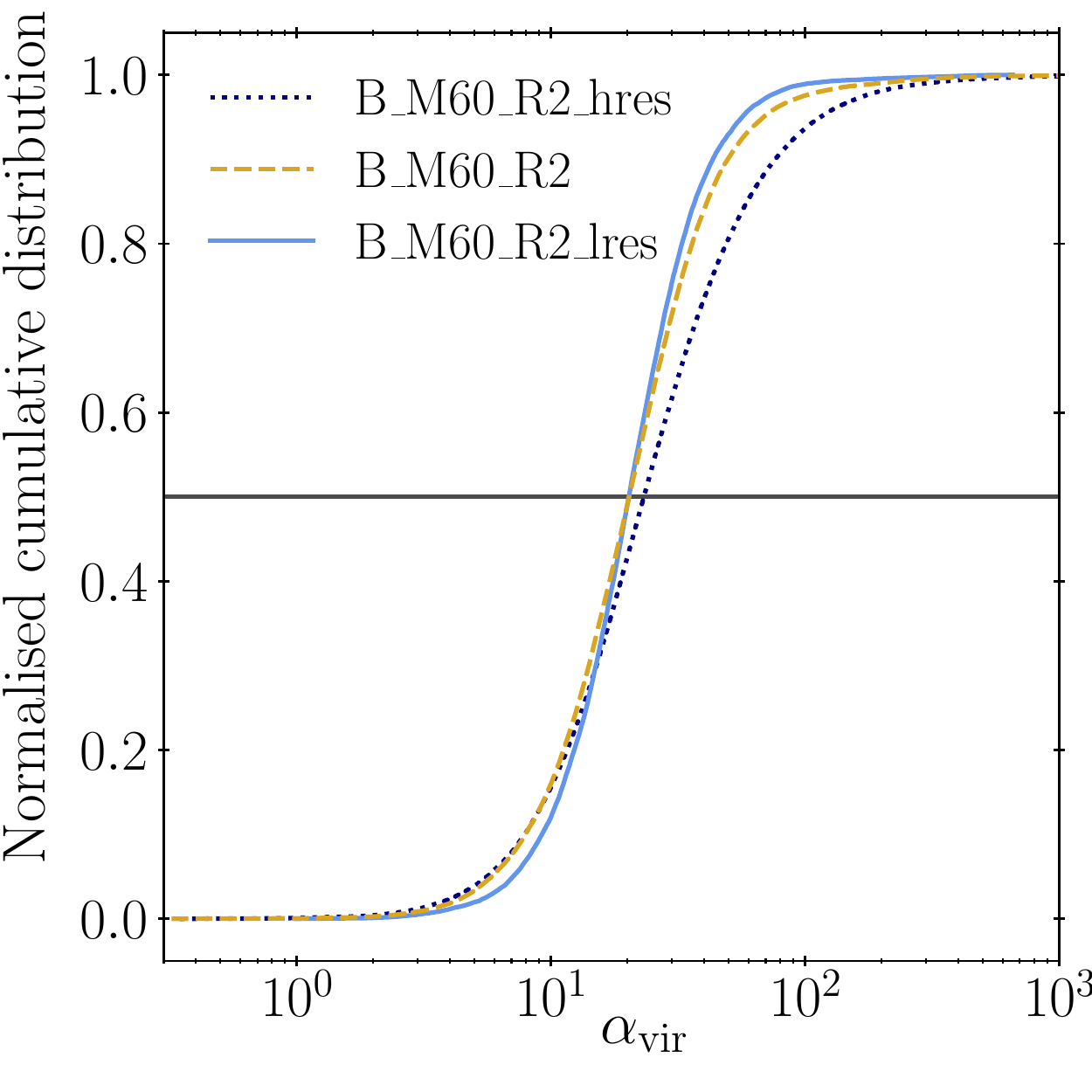}%
    \caption{
    Normalised cumulative distribution of the virial parameter for the different resolutions tested, as indicated by the legend. The horizontal black line indicates the median of each distribution.
    }
    \label{fig:avir_cumulative_distbn}
\end{figure}

\begin{figure}
    \centering
    \includegraphics[width=1.\linewidth]{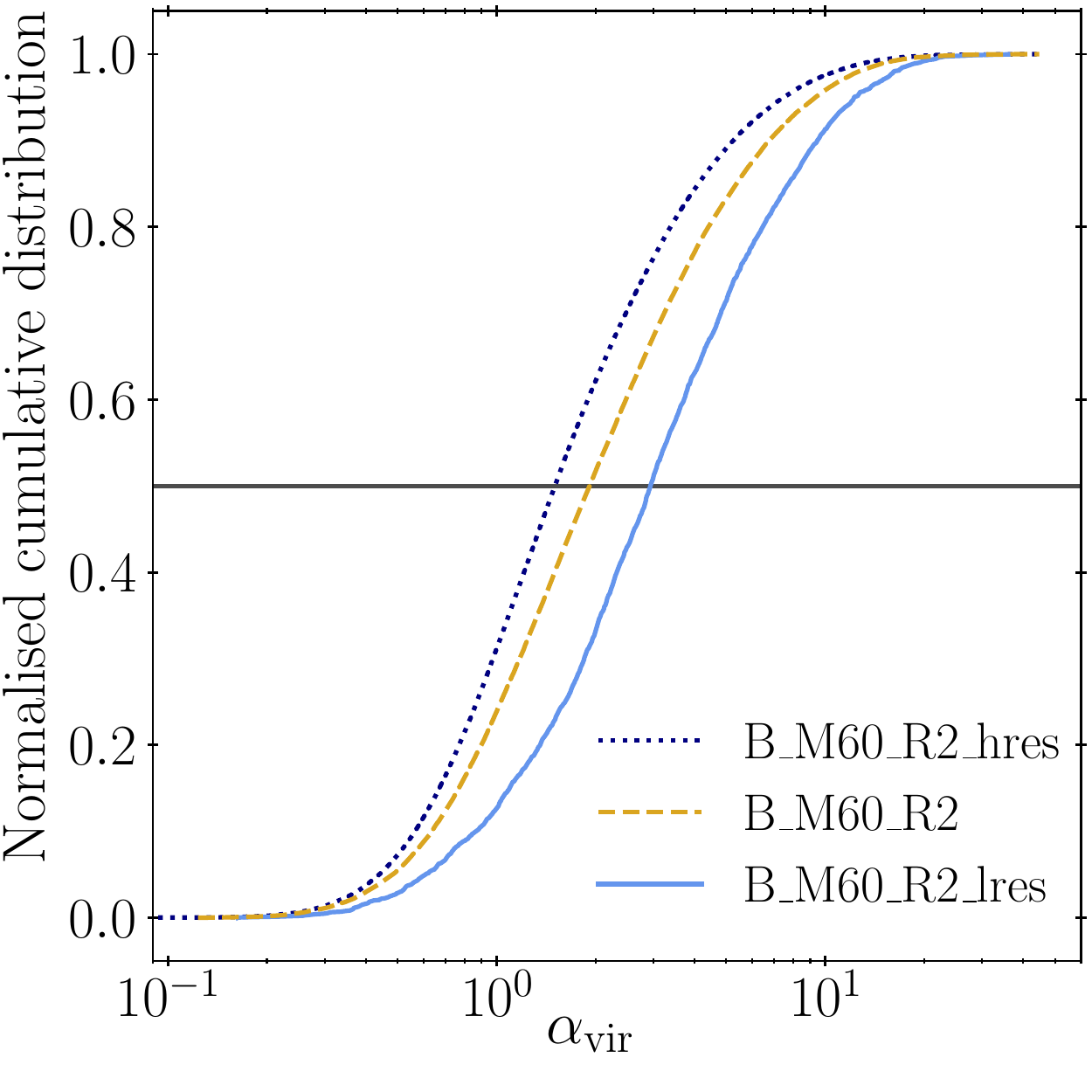}%
    \caption{
    Normalised cumulative distribution of the virial parameter with which stars formed, comparing the three simulations with different resolutions, as indicated by the legend. The horizontal black line indicates the median of each distribution. The median $\avir$ with which stars form lies between $\avir=1.3$ for B\_M60\_R2\_hres and $\avir=2.9$ for B\_M60\_R2\_lres. 
    }
    \label{fig:avir_cumulative_distbn_stars_formed}
\end{figure}

Figure~\ref{fig:avir_cumulative_distbn} shows the cumulative distribution of the virial parameter of all gas cells. This figure illustrates that, despite the order of magnitude difference in resolution, there is very little difference between the distribution of virial parameters in the three simulations. The median of the high-resolution run is slightly higher compared to the other two, but due to the relatively modest dependence of the SFE on the virial parameter (see equation~\ref{eq:eff}), this only makes a difference of a few percent in SFE, demonstrating good convergence. This is further corroborated by Figure~\ref{fig:avir_cumulative_distbn_stars_formed}, which shows the cumulative distributions of the virial parameters from which stars \textit{actually} formed in the three simulations. While the median values vary between 1.3 and 2.9 from the highest to the lowest resolution, this is an acceptable difference considering the factor-of-10 difference in resolution.

Combining the above tests, we conclude that our star formation model reaches satisfactory convergence at the $10^4~\Msun$ resolution that we have used in this study.

\section{Choice of star formation thresholds} \label{app:SFC}
To ascertain the robustness of our sub-grid star formation model, and to ensure that our results are not influenced by the choices of thresholds determining which gas cells are eligible for star formation, we simulate the fiducial bulge galaxy with a range of different density and temperature thresholds for star formation. These thresholds are listed in Table~\ref{tab:ResTestICs}, and include both a higher and lower density threshold by an order of magnitude, as well as a temperature threshold that is higher by a factor of 50. 

\begin{table}
 \begin{tabular}{lccc}
  \hline
  Name & Density threshold [$\ccm$] & Temperature threshold [K] \\
  \hline
  
  B\_M60\_R2 & 1 & $1 \times 10^3$ \\
  B\_M60\_R2\_lDT & 0.1 & $1 \times 10^3$ \\
  B\_M60\_R2\_hDT & 10 & $1 \times 10^3$ \\
  B\_M60\_R2\_hTT & 1 & $5 \times 10^4$ \\

  \hline
  \hline
 \end{tabular}
 \caption{Summary of simulations run to test the influence of star formation thresholds on the dynamics-dependent sub-grid star formation model. All simulations are those of the fiducial bulge ($\Rb = 2~\kpc$ and $\Mb = 0.6\Mstar$). We only vary the listed threshold parameters.}
 \label{tab:ResTestICs}
\end{table}

\begin{figure}
    \centering
    \includegraphics[width=0.95\linewidth]{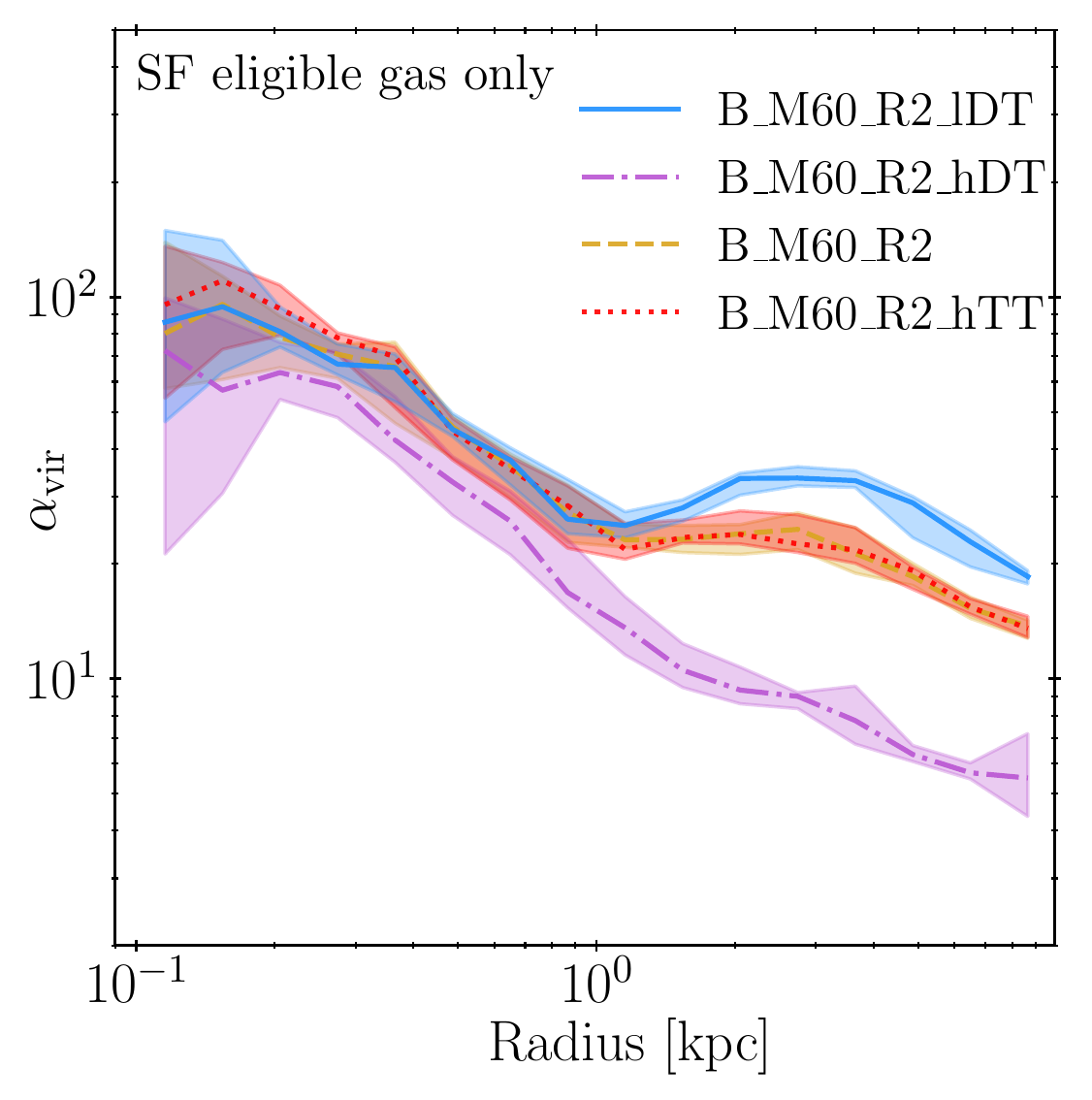}%
    \caption{Time-averaged radial profiles of the virial parameter, comparing simulations with different star formation thresholds, as detailed in Table~\ref{tab:ResTestICs}.
    } 
    \label{fig:radprof_avir_diffSFthresh}
\end{figure}

\begin{figure}
    \centering
    \includegraphics[width=0.95\linewidth]{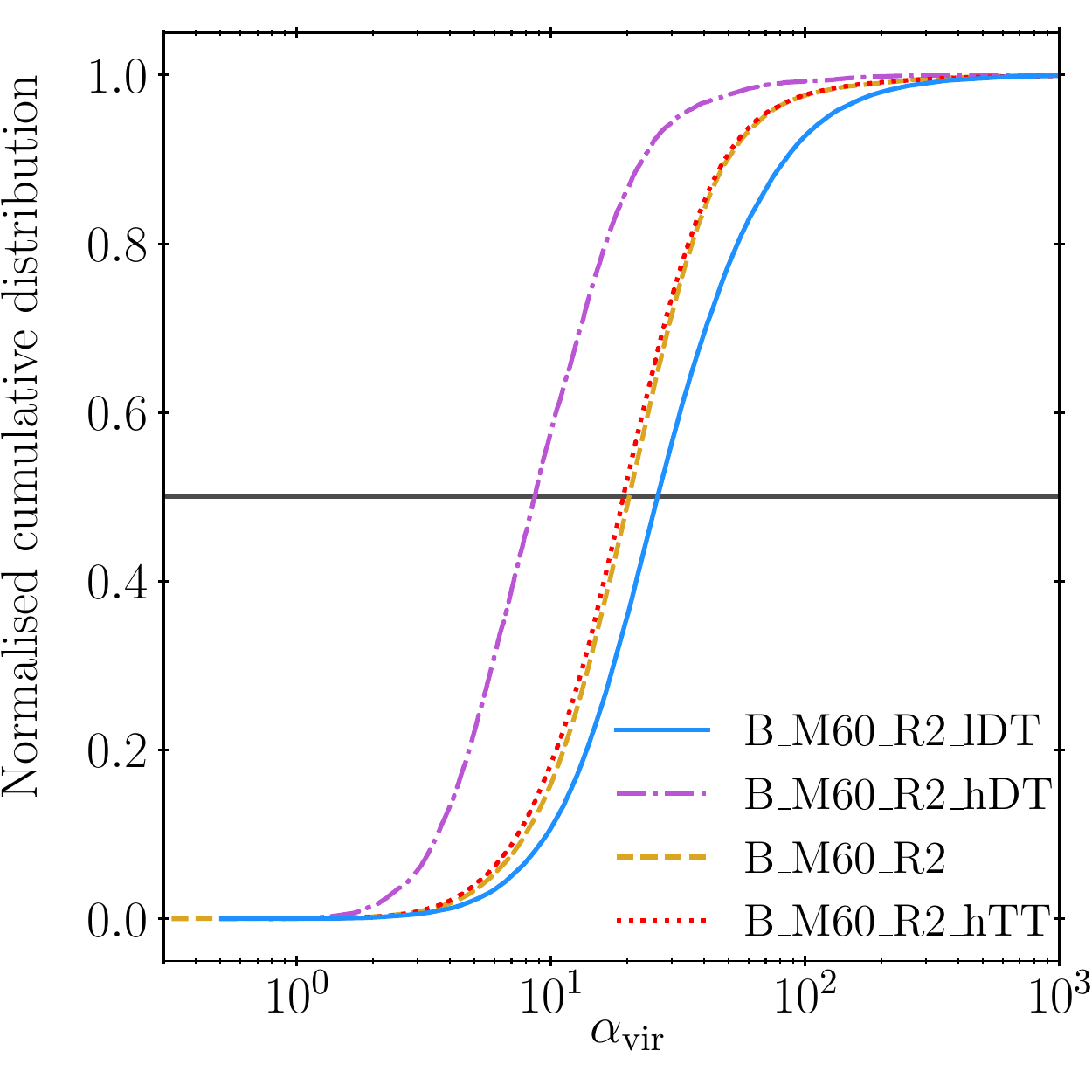}%
    \caption{Normalised cumulative distribution of $\avir$, comparing simulations with different star formation thresholds, as detailed in Table~\ref{tab:ResTestICs}.} 
    \label{fig:avir_cumulative_distbn_diffSFthresh}
\end{figure}

For the gas that is eligible for star formation, Figure~\ref{fig:radprof_avir_diffSFthresh} shows that the gas virial parameters change considerably when using a different density threshold. Higher-density gas is more likely to be self-gravitating and thus have a lower virial parameter. Similarly, including more diffuse gas in the median will push the $\avir$ to higher values. Similar differences are shown in Figure~\ref{fig:avir_cumulative_distbn_diffSFthresh}, which shows the normalised cumulative distributions of the virial parameter in the different simulations.

\begin{figure}
    \centering
    \includegraphics[width=0.95\linewidth]{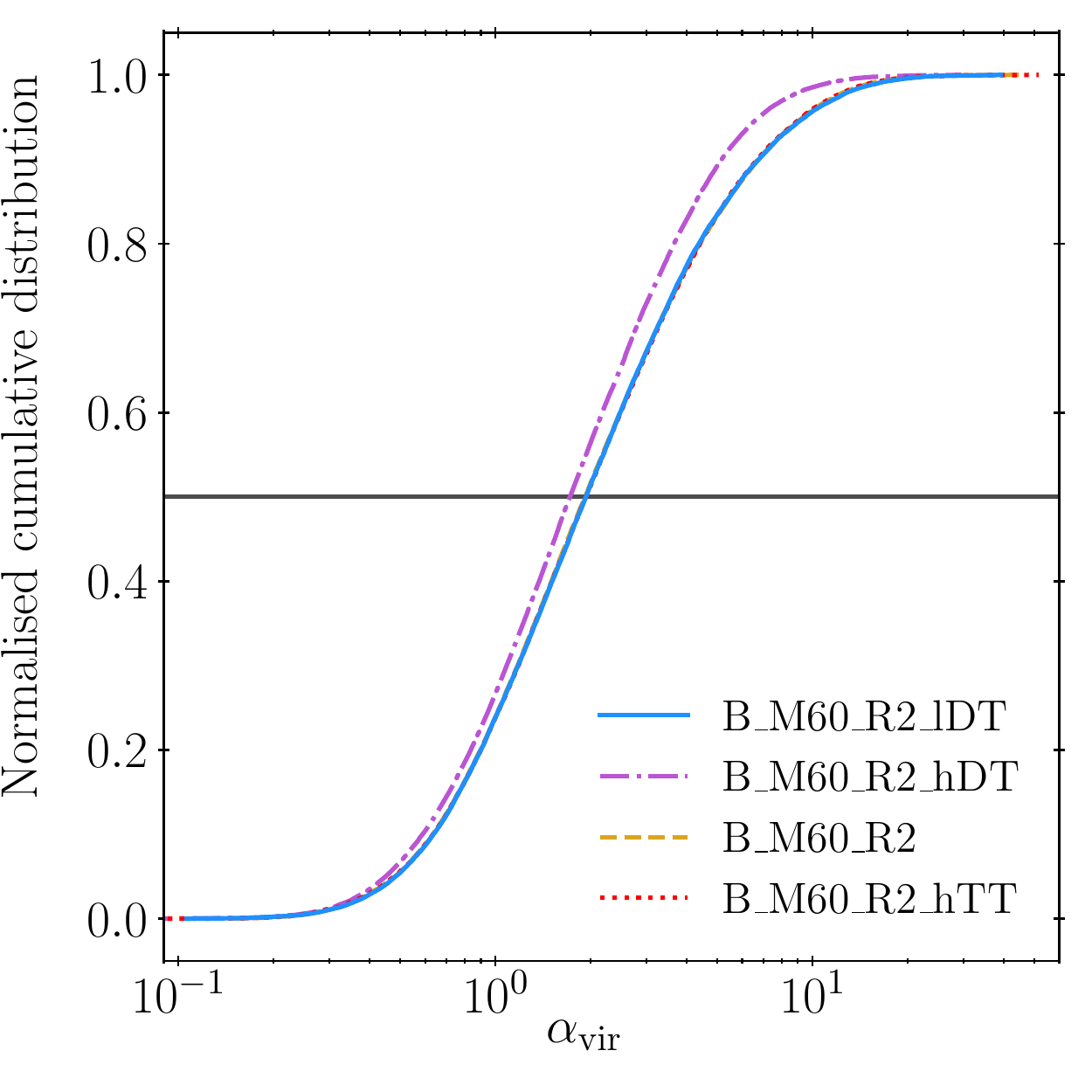}%
    \caption{
    Normalised cumulative distribution of $\avir$ with which stars have formed after $1~\Gyr$, comparing simulations with different star formation thresholds, as detailed in Table~\ref{tab:ResTestICs}.
    } 
    \label{fig:avir_cumulative_distbn_diffSFthresh_stars}
\end{figure}

However, the gas with low virial parameters at which stars form most efficiently represents only a small fraction of the total mass. As a result, the differences in the virial parameter, which mostly manifest themselves at high $\avir$, do not strongly affect the global SFR. This is demonstrated by Figure~\ref{fig:avir_cumulative_distbn_diffSFthresh_stars}, where only the virial parameter in simulation B\_M60\_R2\_hDT shows a small offset, whereas the other three runs agree excellently.

\begin{figure}
    \centering
    \includegraphics[width=0.95\linewidth]{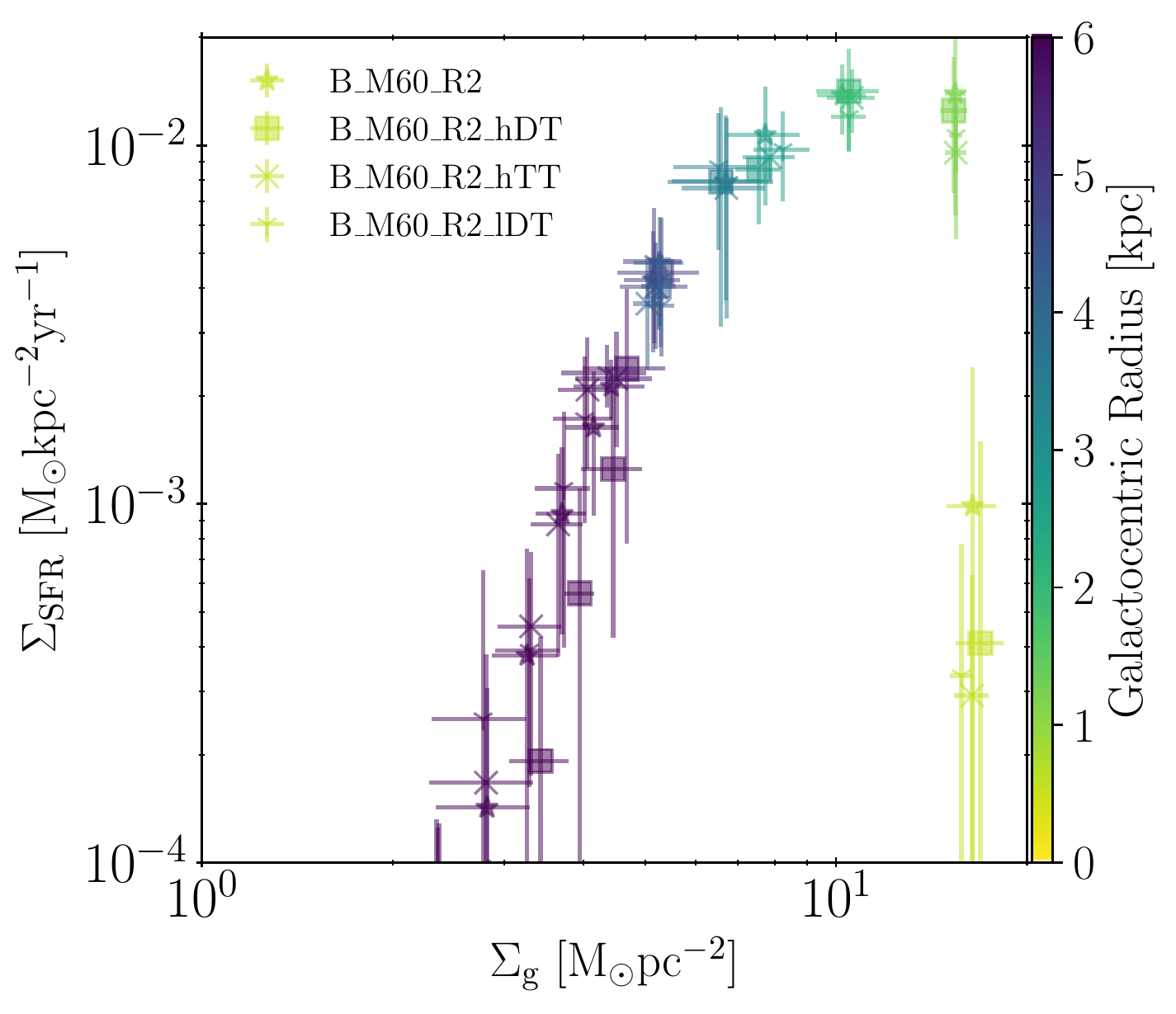}%
    \caption{
    SFR surface density as a function of gas surface density, comparing simulations with different star formation thresholds, as detailed in Table~\ref{tab:ResTestICs}. The colour coding indicates the galactocentric radius.} 
    \label{fig:single_panel_KS_diffSFthresh}
\end{figure}

Finally, Figure~\ref{fig:single_panel_KS_diffSFthresh} shows the star formation relation, generated in the same way as those discussed in Section~\ref{ss:vcSFE} and Section~\ref{ss:vSFE}. All simulations are quantitatively consistent with the fiducial simulation B\_M60\_R2. For all of them, the $\Ssfr$ flattens towards the centre of the galaxy with increasing $\Sg$, with a pronounced suppression in the central radial bin. Any differences agree to within the uncertainties implied by the time variability of the SFR. The insensitivity of the star formation relation to the density and temperature thresholds used is a result of the fact that our dynamics-dependent star formation model self-consistently selects self-gravitating overdensities by increasing their SFE. Irrespectively of the thresholds set, stars form at the roughly same rate, with a median $\avir\sim 1.9$, and follow the same star formation relation.

We conclude that the density and temperature thresholds for star formation used elsewhere in the literature, and which originally have been tuned to work for a constant $\eff$, can also be used with our model. They do not affect our conclusions regarding the influence of the gravitational potential on the SFR of galaxies. As our tests demonstrate, our sub-grid star formation model could also be used without any threshold at all, even if this would imply an increase in computational expense.


\bsp    
\label{lastpage}
\end{document}